\documentclass[11pt,a4paper]{article}
\usepackage{jheppub}

\usepackage{amsmath, amsfonts, amsthm, amssymb, graphicx}
\usepackage[dvipsnames]{xcolor}
\usepackage{epstopdf}
 \usepackage{slashed}
\usepackage{changepage}
 \usepackage{vector}
\usepackage{soul}

\usepackage{pifont}
\def\be{\begin{equation}}
\def\ee{\end{equation}}
\def\ba{\begin{eqnarray}}
\def\ea{\end{eqnarray}}

\def\bdm{\begin{displaymath}}
\def\edm{\end{displaymath}}

\def\bq{\begin{quote}}
\def\eq{\end{quote}}

\def\ltap{\ \raise.3ex\hbox{$<$\kern-.75em\lower1ex\hbox{$\sim$}}\ }
\def\gtap{\ \raise.3ex\hbox{$>$\kern-.75em\lower1ex\hbox{$\sim$}}\ }
\def\gl{\ \raise.5ex\hbox{$>$}\kern-.8em\lower.5ex\hbox{$<$}\ }
\def\roughly#1{\raise.3ex\hbox{$#1$\kern-.75em\lower1ex\hbox{$\sim$}}}

 at 10truept

\newcommand{\beq}{\begin{equation}}
\newcommand{\eeq}{\end{equation}}
\newcommand{\bea}{\begin{eqnarray}}
\newcommand{\eea}{\end{eqnarray}}
\newcommand{\beqa}{\begin{eqnarray}}
\newcommand{\eeqa}{\end{eqnarray}}

\def\be{\begin{equation}}
\def\ee{\end{equation}}

\usepackage[dvipsnames]{xcolor}
\numberwithin{equation}{subsection}

\title{Kerr-like metric in 4D Double Field Theory}

\author[1]{Shunrui Li} 
\author[2,3]{and Yang Liu}

\affiliation[1]{Department of mathematicas and applied mathematics, Beijing University of Chemical Technology, Beijing 100029, China}
\affiliation[2]{School of Physics and Astronomy, University of Nottingham, University Park, Nottingham NG7 2RD, United Kingdom}
\affiliation[3]{Nottingham Centre of Gravity, University of Nottingham, Nottingham NG7 2RD, UK}

\emailAdd{shunruili687@gmail.com}
\emailAdd{yang.liu@nottingham.ac.uk}

\abstract{Double Field Theory suggests that people can view the whole massless NS-NS sector as the gravitational unity. The $O(D,D)$ covariance and the doubled diffeomorphisms determine precisely how the Standard Model as well as a relativistic point particle should couple to the NS-NS sector. The theory also refines the notion of singularity. In \cite{KPS2017}, the authors derive analytically the most general, spherically symmetric, asymptotically flat, static vacuum solution to $D = 4$ Double Field Theory. The solution contains three free parameters and consequently generalizes the Schwarzschild geometry. In this paper, we generalize the metric in \cite{KPS2017} to obtain the 'Kerr-like' metric in $D=4$ double field theory (DFT) in the Einstein frame and string frame. Then we apply 'covariant phase space' approach to study the thermodynamic properties of the metric we have obtained. We explore the first law of black hole thermodynamics and Hawking radiation for this metric carefully. As a special case, the 'Schwarzschild-like' metric in $4D$ double field theory in \cite{KPS2017} can be recovered.} 

\begin{document}
\maketitle

\section{Introduction}
In recent years, one of developments in string theory is double field theory (DFT) \cite{Siegel19931,Siegel19932,D1990}, which describes supergravity in a manifestly duality invariant manner. DFT introduces the dual coordiantes $\tilde{x}_{\mu}$ such that all fields and gauge parameters may depend on the $2D$ coordinates $X^M=(x^{\mu}, \tilde{x}_{\mu})$, which form an $O(D,D)$ vector. In order for the action of generalised diffeomorphisms to give a closed algebra, people need to impose further conditions on the fields and gauge parameters of double field theory. The easiest way is to require that the fields and gauge parameters only depend on at most half of the coordinates, in which case the DFT action can be reduced back to the following action:
\begin{equation} \label{eq:NSNSaction} 
S_{NS-NS} = \int d^D x  \sqrt{\det g} e^{-2\phi} \left(R + 4(\nabla \phi)^2 - \frac{1}{12} H^2 \right), \qquad H_{\mu\nu\rho} \equiv 3 \partial_{[\mu}B_{\nu\rho]},
\end{equation}
where $g_{\mu\nu}$ is the spacetime metric, $B_{\mu\nu}$ 2-form and $\phi$ dilaton. The constraint is called the $section \ condition$ or $strong \ constraint$. Therefore, double field theory provides a natural framework to study possibilities suggested by T-duality which go beyond supergravity. Imposing the section condition implies that all supergravity backgrounds can be viewed as solutions of DFT \cite{ASA and CDAB 2017}.

Black hole (BH) solutions are an excellent setup to test string theory, which is the only consistent theory of quantum gravity so far. The matching between the microscopic states counting for the 5-dimensional BPS black hole solution of the string effective action and the Bekenstein-Hawking entropy is still one of the main successes of superstring theory \cite{SW1996,Z2023}. Furthermore, the non-extremal version of such BH has been studied and some limits of the Bekenstein-Hawking entropy have been matched with the microscopic states counting using U-duality arguments \cite{HMS1996}. Since then, a lot of following work has been done to extend the results of black hole in string theory to other solutions and at higher order in $\alpha'$. The research has two sides. On the one hand, from the macroscopic side, a major development is due to the introduction of the entropy function formalism \cite{S2008}, whose ideas have been used to compute the corrections to the entropy of asymptotically-flat extremal BHs using only their near horizon limit \cite{CDKL20071,CDKL20072,CDKL2008}. On the other hand, from the microscopic side, the corrections to the Cardy formula have been computed in \cite{CM2009,KLL1999,KL2005} finding a perfect match. 

Now we discuss the black hole solutions in double field theory. If we treat the whole closed string massless sector as stringy graviton fields, then double field theory may evolve into Stringy Gravity, i.e. the stringy augmentation of General Relativity \cite{ACP2018}. In \cite{ACP2018}, equipped with an $O(D,D)$ covariant differential geometry beyond Riemann, the authors spell out the definition of the energy–momentum tensor in Stringy Gravity and derive the corresponding on-shell conservation law from doubled general covariance. Equating it with the recently identified stringy Einstein curvature tensor, all the equations of motion of the closed string massless sector can be unified into a single expression, $G_{AB}= 8\pi G T_{AB}$, which is the so-called Einstein Double Field Equations. In \cite{KPS2017,ACP2018}, the authors study the most general $D = 4$ static, asymptotically flat, spherically symmetric, ‘regular’ solution, sourced by the stringy energy–momentum tensor which is nontrivial only up to a finite radius from the center. Outside this radius, the solution matches the known vacuum geometry which has four constant parameters. Furthermore, the authors express these as volume integrals of the interior stringy energy–momentum tensor and discuss the related energy conditions of the black hole metric \cite{ACP2018}.

The metric in \cite{KPS2017,ACP2018} is 'Schwarzschild-like' metric in $4D$ DFT. In this paper, we will generalize this metric to rotating case, namely, 'Kerr-like' metric. The main result of our investigation has two parts. First, we obtain the Kerr-like metric in $4D$ double field theory in the Einstein frame and string frame. We mimic the derivation method of Kerr metric in General Relativity and obtain the Kerr-like metric in $D=4$ DFT. We do the complexification for the physical quantities in this metric. We discuss the conditions which dilaton and $H$-flux should satisfy as well. Second, we study the black hole theormodynamics and Hawking radiation of the Kerr-like metric we have obtained. We can make use of the 'covariant phase space' approach \cite{V.L and R.M.W 1994,R.M.W 1993,J.L and R.M.W 1990}. In this approach the first law of black hole thermodynamics is re-expressed in a 'differential' form, as the vanishing of the exterior derivative of a certain $(D-2)$-form constructed out of the fields and their variations. Then Stokes' theorem sets the integral of this form on a horizon cross-section equal to the integral on a sphere at infinity, recovering the usual, integrated form of the first law of black hole thermodynamics \cite{ASA and CDAB 2017}.

The rest of this paper is organised as follows.  In section 2, we review the main results of background we use in this paper. In section 3, we derive the Kerr-like metric in $4D$ double field theory. In section 4, we discuss the thermodynamic properties of the Kerr-like metric which is obtained in section 3. In section 5, we conclude our results and discuss some remaining concerns.

\section{Basics}
\subsection{Review of double field theory}
Double field theory (DFT) generalizes the spacetime action, which can possess T-duality on the level of component fields manifestly \cite{WY2014}. Early works can see \cite{T1990,D1990}. Due to the equivalence of spacetime momenta and winding numbers in the string spectra \cite{WY2014}, a set of conjugated coordinates $\tilde{x}^i$, which is conjugated to winding numbers, can be introduced naturally \cite{WY2014}. These conjugated coordinates are treated on the same footing as the usual coordinates $x_i$. Then the dimension of spacetime is doubled from $D$ to $D+D$ \cite{WY2014}.
The action of DFT unifies the metric $g_{ij}$, the two-form $b_{ij}$ and the dilaton $\phi$ by rewriting these fields in an $O(D,D)$ covariant way.
The group $O(D,D)$ itself is defined to the set of transformations perserving the $O(D,D)$ structure\\ 
\begin{equation} \label{eq:eta} 
    \eta_{MN}=\left[\begin{array}{cc}
      0   & I_D \\
      I_D   & 0
    \end{array}\right],
\end{equation}
while the generalized metric $\mathcal{H}_{MN}$ is given by 
\begin{equation}\label{eq:H}
\mathcal{H}_{MN}
=
\begin{bmatrix}
g^{ij} & -g^{ik} B_{kj}\\
B_{ik}g^{kj} & g_{ij}- B_{ik} g^{kl} B_{lj}
\end{bmatrix}.
\end{equation}
The generalised metric is symmetric and constrained to satisfy $\mathcal{H}_M^N \mathcal{H}_N^P=\delta_M^P$, which implies that it parameterises the coset $O(D,D)/O(1,D-1) \times O(1,D-1)$. \\
If there is no dependence on the conjugated coordinates, the DFT can be reduced to the supergravity action \cite{B2019}. The action is given by 
\begin{equation}\label{eq:supergravity action}
S =\frac{1}{16\pi G_{DFT}} \int dx d\tilde{x} e^{-2d} \mathcal{R}, 
\end{equation}
where $d$ is the shifted dilaton, which contains the determinant of the metric and the usual dilaton $\phi$ both \cite{B2019,HZ20101}, i.e.,
\begin{equation}\label{eq:usual dilaton} 
e^{-2d} = \sqrt{-g} e^{-2\phi}, 
\end{equation}
and generalised Ricci scalar $\mathcal{R}$ is
\begin{align}  \label{RS_R}
    \mathcal{R}=& \frac{1}{8} \mathcal{H}^{MN} \partial_{M} \mathcal{H}^{KL} \partial_{N} \mathcal{H}_{KL} - \frac{1}{2} \mathcal{H}^{MN} \partial_{N} \mathcal{H}^{KL} \partial_{L} \mathcal{H}_{MK} - \partial_{M} d \partial_{N} \mathcal{H}^{MN}+ \\ \nonumber
    &4 \mathcal{H}^{MN} \partial_{M} \partial_{N} d -4H^{MN}\partial_Md \partial_Nd+4\partial_M H^{MN}\partial_N d.
\end{align}
The gauge fields $A_{\mu}^M$ and $B_{\mu \nu}$ constitute the "tensor hierarchy" of the split theory.
The field strengths are
\begin{equation} \label{F}
    \mathcal{F}^M_{\mu \nu}=\partial_{\mu}A^M_{\nu}-\partial_{\nu}A^M_{\mu}-[A_{\mu},A_{\nu}]^M_C+\partial^MB_{\mu \nu},
\end{equation}
\begin{equation} \label{H}
    \mathcal{H}_{\mu \nu \rho}=3D_{[\mu}B_{\nu \rho]}-3A_{[\mu}^N\partial_{\nu}A_{\rho]N}+A_{[\mu|N|}[A_{\nu},A_{\rho]}]_C^N,
\end{equation}
and the double field internal index $C_{\mu \nu}$ can be defined by
\begin{equation} \label{C mu nu}
    C_{\mu \nu} \equiv -B_{\mu \nu}+\frac{1}{2}A_{\mu}^M A_{\nu M}.
\end{equation}
In the above, we have introduced the derivative $D_{\mu} = \partial_{\mu}-\mathcal{L}_{A_{\mu}}$, which is covariant under generalised diffeomorphisms, as explained in \cite{OH and HS 2013}.

\subsection{Spherical solutions in D=4 double field theory}
The massless spectrum of any of closed string theories has a common sector of  the NS-NS field: the spacetime metric $g_{\mu\nu}$, 2-form $B_{\mu\nu}$ amd dilaton $\phi$, the low energy of these field is
\begin{equation} \label{eq: NS-NS action}
    S_{NS-NS}=\int d^Dx \sqrt{g} e^{-2\phi} (R+4(\nabla \phi )^2-\frac{1}{12}H_{\mu \nu \omega}H^{\mu \nu \omega}),
\end{equation}
where $H_{\mu \nu \omega}\equiv 3\partial_{[\mu}B_{\nu \omega]}$. 
We can obtain generalised Einstein equations for pure double field theory by variational method,
\begin{equation} \label{eq: generalsed Einstein equation 1}
    R_{\mu \nu} +2\nabla_{\mu} (\partial_{\nu} \phi)-\frac{1}{4}H_{\mu \rho \sigma}H_{\nu}^{\rho \sigma} = 0,
\end{equation}
\begin{equation} \label{eq: generalsed Einstein equation 2}
    d\star (e^{2\phi}H_{(3)})=0,
\end{equation}
\begin{equation} \label{eq: generalsed Einstein equation 3}
    R+4\Box \phi -4 \partial_{\mu}\phi \partial^{\mu}\phi-\frac{1}{12}H_{\mu \rho \sigma}H^{\nu \rho \sigma} = 0.
\end{equation}
More details can be found in \cite{JHP2019, KPS2017}.

We review the most general form of the static, asymptotically flat and spherically symmetric vacuum solutions to $D=4$ double field theory briefly \cite{KPS2017, ACP2018}. Without loss of generality, the metric for the string frame can be assumed as 
\begin{equation}\label{eq:metric under string frame (inital)}
ds^2 = e^{2\phi(r)} [-A(r)dt^2 + A(r)^{-1} dr^2 + A(r)^{-1} C(r)d\Omega^2], 
\end{equation}
where
\begin{equation}\label{eq:metric under string frame omega}
 d\Omega^2= d\theta^2 + \sin^2 \theta d\varphi^2.
\end{equation}
It is worthwhile to note that our string frame metric ansatz takes the product form of the dilaton factor, $e^{2\phi}$, times the Einstein frame metric.\\
If the spacetime is asymptotically 'flat', then the metric should satisfy the following three boundary conditions \cite{KPS2017}:
\begin{equation}\label{eq:limitation A(r)}
\lim_{r \rightarrow \infty} A(r) =1,
\end{equation} 
\begin{equation}\label{eq:limatation r-2C(r)}
\lim_{r \rightarrow \infty} r^{-2} C(r) =1,
\end{equation}
\begin{equation}\label{eq:limiation phi(r)}
\lim_{r \rightarrow \infty} \phi(r) =0.
\end{equation}
From the asymptotic ‘smoothness’, the metric should satisfy:
\begin{equation}\label{eq:limitation A(r)' A(r)''}
\lim_{r \rightarrow \infty} A'(r) = \lim_{r \rightarrow \infty} A''(r) = 0,
\end{equation} 
\begin{equation}\label{eq:eq:limitation r-1C(r)' C(r)''}
\lim_{r \rightarrow \infty} r^{-1} C'(r) = \lim_{r \rightarrow \infty} C''(r) =2,
\end{equation}
\begin{equation}\label{eq:limitation phi(r)' phi(r)''}
\lim_{r \rightarrow \infty} \phi'(r) = \lim_{r \rightarrow \infty} \phi''(r) =0.
\end{equation}
We write the $B$-field using the form notation \cite{KPS2017}
\begin{equation}\label{eq:B2}
B_{(2)}= \frac{1}{2} B_{\mu\nu} dx^{\mu} \wedge dx^{\nu} = B(r) \cos \theta dr \wedge d\varphi + h \cos \theta dt \wedge d\varphi.
\end{equation}
The $H$-flux, which takes the most general spherically symmetric form, can be written as \cite{KPS2017}
\begin{equation}\label{eq:H3}
H_{(3)}= \frac{1}{3!} H_{\lambda\mu\nu} dx^{\lambda} \wedge dx^{\mu} \wedge dx^{\nu} = B(r) \sin \theta dr \wedge d\theta \wedge d\varphi + h \sin \theta dt \wedge d\theta \wedge d\varphi,
\end{equation}
which is closed for constant $h$ \cite{KPS2017}. As a result, with four constants $a$, $b$, $c$, $h$ and \cite{KPS2017}
\begin{equation}\label{eq:c+}
c_{+} = c + \frac{1}{2} \sqrt{a^2 + b^2},
\end{equation}
\begin{equation}\label{eq:c-}
c_{-} = c - \frac{1}{2} \sqrt{a^2 + b^2},
\end{equation}
\begin{equation}\label{eq:gamma+-}
\gamma_{\pm} = \frac{1}{2} (1 \pm \sqrt{1-h^2/b^2}),
\end{equation}
the metric can be given \cite{KPS2017, ACP2018}:
\begin{equation}\label{eq:e 2 phi}
e^{2\phi} = \gamma_{+} \left(\frac{r- \alpha}{r+\beta}\right)^{\frac{b}{\sqrt{a^2 + b^2}}} +  \gamma_{-} \left(\frac{r+ \beta}{r-\alpha}\right)^{\frac{b}{\sqrt{a^2 + b^2}}}.
\end{equation}
It can be proved that the asymptotic flatness is inconsistent with the magnetic $H$-flux, therefore we should set $B(r)=0$ \cite{KPS2017}. Then we have
\begin{equation} \label{B}
    B_{(2)}=h \cos \theta dt \wedge d\varphi,
\end{equation}
\begin{equation} \label{H}
   H_{(3)}= h \sin \theta dt \wedge d\theta \wedge d\varphi.
\end{equation}
\begin{equation}\label{eq:solution}
ds^2= e^{2\phi} [- \left(\frac{r- \alpha}{r+\beta}\right)^{\frac{a}{\sqrt{a^2 + b^2}}} dt^2 +  \left(\frac{r+ \beta}{r-\alpha}\right)^{\frac{a}{\sqrt{a^2 + b^2}}} \times \{dr^2 + (r-\alpha)(r+\beta) d\Omega^2\} ],
\end{equation}
where
\begin{equation} \label{alpha}
\alpha=\frac{a}{a+b} \sqrt{a^2+b^2},
\end{equation}
and
\begin{equation} \label{beta}
\beta=\frac{b}{a+b} \sqrt{a^2+b^2}.
\end{equation}

\subsection{Noether charge and the first law of black hole in double field theory}
For the black hole in general relativity, Robert M.Wald, Joohan Lee and Vivek Lyer verified that black hole entropy is Noether charge by covariant phase space formalism \cite{R.M.W 1993, J.L and R.M.W 1990, V.L and R.M.W 1994}.\\
The Noether charge $(n-2)$-from can always be expressed in the form
\begin{equation} \label{Noether 1}
    \mathcal{Q}= \textbf{W}_c(\phi)\xi^c+\textbf{X}^{cd}\nabla_{[c}\xi_{d]}+d\textbf{Z}(\phi,\xi),
\end{equation}
where $\textbf{W}_c, \textbf{X}^{ab}, \textbf{Y}$ and $\textbf{Z}$ are covariant quantities which are locally constructed from the indicated fields and their derivatives (with $\textbf{Y}$ linear in $\mathcal{L}_{\xi} \phi$ and \textbf{Z} linear in $\xi$).\\
For DFT, it is more convenient to use the variation of action
\begin{equation} \label{variation of action in DFT}
    \delta S_{DFT}=\int e^{-2d}(-2\delta d \mathcal{R}+\delta H^{MN} \mathcal{R}_{MN})+\int \partial_M(e^{-2d} \Theta^M[H,dt;\delta H, \delta d],
\end{equation}
where
\begin{align} \label{ThetaM}
    \Theta^M=&\delta H^{QP}(\frac{1}{4}H^{MR}\partial_R H_{QP}-\frac{1}{2}H^{MR}\partial_QH_{RP} -\frac{1}{2}H^R_Q\partial_RH^M_P)+ \\ \nonumber
    &2\delta H^{MP}\partial_Pd-\partial_P\delta H^{MP}+4\partial_P(\delta d)H^{MP}.
\end{align}
We consider the varition of the DFT action under a generalised diffeomorphism with parameter $\Lambda^M$, 
\begin{equation} \label{variation of action in dft guage}
    \delta_{\Lambda}S_{DFT}=\int \partial_M (\Lambda^M e^{-2d} \mathcal{R}).
\end{equation}
By comparing the two variations \eqref{variation of action in DFT} and \eqref{variation of action in dft guage}, we can get a boundary term 
\begin{equation} \label{J}
    \mathcal{J}=e^{-2d}(\Theta[H,d;\mathcal{L}_{\Lambda}H, \mathcal{L}_{\Lambda}d]-\Lambda^M \mathcal{R}),
\end{equation}
which is divergence-free whenever $(d,H)$ are on-shell
\begin{equation} \label{close}
    \mathcal{R}=0=\mathcal{R}_{MN} \rightarrow \partial_M \mathcal{J}^M=0,
\end{equation}
and an anstisymmetric $\mathcal{J}^{MN}$ satisfies
\begin{equation} \label{anstisymmetric J}
    \mathcal{J}^M=\partial_N \mathcal{J}^{MN}.
\end{equation}
In DFT we have 
\begin{equation} \label{slashed delta Q}
    \slashed{\delta} Q_{\Lambda}=\Omega[d,H;(\delta d, \delta H),(\mathcal{L}_{\Lambda}d,\mathcal{L}_{\Lambda}H)].
\end{equation}
We will consider it as a definition of the infinitesimal Noether charge $\slashed{\delta} Q_{\Lambda}$ associated to $\Lambda^M$ and more details can be found in \cite{R.M.W 1993, J.L and R.M.W 1990, V.L and R.M.W 1994}.\\
Additionally, we can write symplectic form on the right hand of \eqref{slashed delta Q} as
\begin{equation} \label{symplectic form}
    \Omega[d,H;(\delta_1 d, \delta_1 H),(\delta_2 d, \delta_2 H)]=\delta_1 [e^{-2d}\Theta^M(H,d;\delta_2H, \delta_2d)]-\delta_2 [e^{-2d}\Theta^M(H,d;\delta_1H, \delta_1d)].
\end{equation}
For double field theory, we also have
\begin{equation} \label{mathcal J}
	\mathcal{J}^{\mu \nu}=e^{-2 \hat{d}}(-2g^{\rho[\mu}\nabla_{\rho}(\xi')^{\nu ]}-\widetilde{\lambda}_pH^{\mu \nu \rho}-\widetilde{\Lambda}^M H_{MN}\mathcal{F}^{\mu \nu N})\epsilon_{\mu\nu},
\end{equation}
where $\xi'$ is unit surface gravity Killing vector, which is defined from Killing vector $\xi$ 
\begin{equation} \label{xi'}
    \xi'=\frac{1}{\kappa}\xi.
\end{equation}
Moreover, in this equation, $\kappa$ is also defined by $\xi$
\begin{equation} \label{kappa}
    \partial_{\mu}(\xi^{\sigma}\xi^{\nu}g_{\sigma \nu})=(-2\kappa)\xi_{\mu},
\end{equation}
where the Killing vector is defined by
\begin{equation} \label{xi}
    \xi^{\mu}=\frac{\partial}{\partial t}.
\end{equation}
Furthermore, on the bifurcation, in \eqref{mathcal J}, we have
\begin{equation} \label{bifurcation}
	\widetilde{\lambda_p} \equiv -(\xi')^{\rho}C_{\rho \mu} \qquad \widetilde{\Lambda}^M \equiv (\xi')^{\rho}A_{\rho}^M.
\end{equation}
The existence of a Noether charge $Q_{\Lambda}$ whose variation equals the right hand of \eqref{symplectic form} is equivalent to the existence of the 'boundary term' $B^M$ such that
\begin{equation} \label{boundary term}
    \delta \int_{\partial C(\infty)} e^{-2d}(2B^M \Lambda^P) \epsilon_{MP}=\int_{\partial C(\infty)} e^{-2d}(2\Theta^{[M} \Lambda^{P]}) \epsilon_{MP}=\int_C \partial_P(2e^{-2d}\Theta^{[M} \Lambda^{P]}) \epsilon_{MP},
\end{equation}
where the $\epsilon$ are the normal and binormal to the codimension 1 “Cauchy surface” $C$ and its boundary at infinity $\partial C(\infty)$ respectively.\\
The Noether charge is
\begin{equation} \label{noether dft}
    Q_{\Lambda}=\int_C \mathcal{J}^M \epsilon_M -\int_{\partial C(\infty)} e^{-2d}(2B^M \Lambda^P)\epsilon_{MP}=\int_{\partial C(\infty)}(\mathcal{J}^M-e^{-2d}2B^M \Lambda^P)\epsilon_{MP}. 
\end{equation}
Let us assume that we have a background for which there exists a horizon specified by $R=R_0$ for a radial coordinate $R$. Then if we integrate against the codimension 1 “Cauchy surface” $C$ given by $t = t_0$, so we can define mass as the Noether charge
\begin{equation} \label{mass}
    M\equiv Q_{\xi}=\frac{1}{16\pi G_{DFT}} \int_{\partial C(\infty)} d^{d-2}x d^{2n}X(\mathcal{J}^{\mu \nu}-e^{-2d}2B^{\mu}\xi^{\nu})\epsilon_{\mu \nu},
\end{equation}
where $n=D-d$. As a result, we can get the first law of black hole thermdynamics
\begin{equation} \label{first law}
    \delta M=\frac{\kappa}{2\pi}\delta S+\slashed{\delta}W,
\end{equation}
where $S$ is entropy, 
\begin{equation} \label{entropy}
    S \equiv 2\pi Q_{hor,\xi'}=\frac{1}{4G_{DFT}}\int_{t=t_0, R=R_0}d^{d-2}xd^{2n}Xe^{-2d}\sqrt{|detg_{d-2}|},
\end{equation}
and $\slashed{\delta}W$ is $O(n,n)$ invariant thermodynamic work contribution,
\begin{equation} \label{slashed delta W}
    \slashed{\delta}W \equiv -\frac{1}{16\pi G_{DFT}}\int_{t=t_0, R=R_0} d^{d-2}xd^{2n}X[\tilde{\lambda}_{\rho}\delta(e^{-2d} \mathcal{H}^{\mu \nu \rho}+\tilde{\Lambda}^M \delta(e^{-2d}H_{MN} \mathcal{F}^{\mu \nu N})]\epsilon_{\mu \nu}.
\end{equation}
Extending \eqref{first law} to the case where the external metric $g_{\mu\nu}$ is only stationary, rather than static, is a straightforward task. However, such an extension necessitates assuming — or proving — a form of horizon rigidity theorem, relevant within the context of the split parametrization elucidated in \cite{OH and HS 2013}. This theorem would ensure the existence of a certain number of commuting Killing vector fields $\partial/\partial\varphi^I$ so that $\xi=\partial/\partial t+\Omega^I\partial/\partial\varphi^I$; the left-hand side of \eqref{first law} would then be replaced by $\delta M-\Omega^I\delta J_I$. Therefore, the first law of black hole thermodynamics for stationary metric is
\begin{equation} \label{first law 2}
    \delta M'=\frac{\kappa}{2\pi}\delta S+\slashed{\delta}W,
\end{equation}
where
\begin{equation} \label{delta W'}
    \delta M'=\delta M- \Omega^I\delta J_I.
\end{equation}
More details of Noether charge and black hole thermodynamics in DFT can be found in \cite{ASA and CDAB 2017}.

\section{Derivation of Kerr-like metric}

In this section, we present the Kerr-like metric in the context of DFT. Initially, our focus lies on deriving the metric for the Einstein frame, excluding considerations related to dilation. We propose a more generalized approach for deriving the metric, utilizing the Schwarzschild-like metric to derive the Kerr-like metric. Subsequently, the metric under this specified condition is acquired through a sequence of transformations. 
Finally, we discuss the conditions which dilaton and $H$-flux should satisfy.

\subsection{General derivation method}
In this subsection, we present a more general derivation method for the Kerr-like metric within the Einstein frame. Initially, we employ transformations aimed at eliminating the $g_{rr}$ term from the metric, while simultaneously transitioning the coordinate system from $(t, r, \theta, \varphi)$ to $(u, r, \theta, \varphi)$. Subsequently, leveraging contravariant coordinates, we embark on a series of transformations, some of which need to do complexification. Upon acquiring the inverse metric, we proceed to obtain the metric in $(u, r, \theta, \varphi)$ coordinates by computing the inverse matrix. Finally, we revert to the original coordinates $(t, r, \theta, \varphi)$ to deduce the Kerr-like metric within the Einstein frame.\\
The Schwarzschild-like metric in $4D$ DFT is given by \eqref{eq:metric under string frame (inital)}. Now,we generalize the spherically symmetric metric to Kerr-like metric. In the following section we will obtain the Kerr-like metric in the Einstein metric. Therefore, $e^{2\phi(r)}$ term should be neglected. First, we can rewrite the angular part of metric \eqref{eq:metric under string frame omega} in a general form 
\begin{equation}\label{new omega}
    d\Omega^2=f_1^2(\theta)d\theta^2+f_2^2(\theta)d\varphi^2.
\end{equation}
We introduce a new variable $u$, which is given by
\begin{equation} \label{du general}
    du=dt-A^{-1}(r)dr.
\end{equation}
Then the metric can be rewritten as
\begin{equation} \label{du ds general}
 ds^2=-A(r)du^2+2dudr+A^{-1}(r)C(r)d{\Omega}^2.
\end{equation}
According to Appendix A, the metric can now be expressed in terms of a null tetrad $\{l^{\mu},n^{\mu},m^{\mu},\bar{m}^{\mu}\}$. The vectors $l^{\mu},n^{\mu}$ are real while $m^{\mu},\bar{m}^{\mu}$ are complex conjugates; the only non-vanishing scalar products between tetrad vectors are: $l^{\mu}n_{\mu}=1$ and $n^{\mu}n_{\mu}=0$. \\ 
As a result, we have:
\begin{equation} \label{g up l n m}
    g^{\mu \nu}=-l^{\mu}n^{\nu}-n^{\mu}l^{\nu}+m^{\mu}\bar{m}^{\nu}+\bar{m}^{\mu}m^{\nu}.
\end{equation}
and
\begin{equation} \label{l generl}
    l^{\mu}=\delta_1^{\mu}, 
\end{equation}
\begin{equation} \label{n general}
    n^{\mu}=\delta_0^{\mu}+\frac{1}{2}A(r)\delta_1^{\mu},
\end{equation}
\begin{equation} \label{m genral}
    m^{\mu}=\frac{1}{\sqrt{2A^{-1}(r)C(r)}}\left(\frac{1}{f_1}\delta_2^{\mu}+\frac{i}{f_2}\delta_3^{\mu}\right),
\end{equation}
\begin{equation} \label{bar m general}
    \bar{m}^{\mu}=\frac{1}{\sqrt{2A^{-1}(r)C(r)}}\left(\frac{1}{f_1}\delta_2^{\mu}-\frac{i}{f_2}\delta_2^{\mu}\right),
\end{equation}
where the $\delta$'s are the unit vectors of cotangent space. \\
Then we introduce two complex transformations
\begin{equation} \label{transformation }
    r'=r+i\psi P(\theta) \: \qquad \text{and} \qquad \: u'=u+i\psi W(\theta),
\end{equation}
where we have defined rotation parameter
\begin{equation} \label{psi}
    \psi=J/M,
\end{equation}
where $J$ is angular momentum.\\
Taking the derivative of \eqref{transformation }, we can get:
\begin{equation} \label{derivative transformation }
dr'=dr+i \psi P'd\theta \: \qquad \text{and} \qquad \: du'=du+i \psi W'd\theta, 
\end{equation}
where $P'$ and $W'$ are the derivatives with respect to $\theta$.\\ 
In fact, \eqref{derivative transformation } can be rewritten as a matrix 
\begin{equation} \label{du dr .. transformation}
    \left[ \begin{array}{c}
        du' \\
        dr' \\
       d\theta' \\
       d\varphi'
    \end{array}
    \right]=
    \left[ \begin{array}{cccc}
        1 & 0 & i \psi P' & 0\\
        0 & 1 & i \psi W' & 0\\
        0 & 0 & 1 & 0\\
        0 & 0 & 0 & 1
    \end{array}
    \right]
    \left[\begin{array}{c}
         du \\
         dr \\
         d\theta \\
         d\varphi
    \end{array}
    \right].
\end{equation}
After applying transformation \eqref{du dr .. transformation}, these unit vectors can be written as:
\begin{equation}
    \delta_0^{\mu}=\delta'^{\mu}_0,
\end{equation}
\begin{equation}
    \delta_1^{\mu}=\delta'^{\mu}_1,
\end{equation}
\begin{equation}
\delta_2^{\mu}=\delta'^{\mu}_2+i\psi P'\delta'^{\mu}_1+i\psi W'\delta'^{\mu}_0, 
\end{equation}
\begin{equation}
    \delta_3^{\mu}=\delta'^{\mu}_3.
\end{equation}
Then we can rewrite $l^{\mu}, n^{\mu}, m^{\mu} $ as:
\begin{equation} \label{general new l}
    l^{\mu}=\delta_1^{\mu},
\end{equation}
\begin{equation} \label{general new n}
    n^{\mu}=\delta'^{\mu}_0+\frac{1}{2}\hat{A}(r)\delta'^{\mu}_1, 
\end{equation}
\begin{equation} \label{general new m}
    m^{\mu}=\frac{1}{\sqrt{2 \widehat{(A^{-1}C)}(r',\bar{r'})}}\left(\frac{1}{f_1}\delta'^{\mu}_2+\frac{i\psi P'}{f_1}\delta'^{\mu}_1+\frac{i\psi W'}{f_1}\delta'^{\mu}_0+\frac{i}{f_2}\delta'^{\mu}_3\right),
\end{equation}
where $\hat{A}(r)$ and $\widehat{A^{-1}C}(r',\bar{r'})$ are the complexification of $A$ and $A^{-1}(r)C(r)$. In subsection 3.2, we will focus on complexification.
The function $\hat{A}(r)$ is obtained by preserving the norm complex continuation for $A(r)$, namely, 
\begin{equation}
    \hat{A} (r)=A(r',\bar{r'}),
\end{equation}
In Kerr-Newman metric, $A=1-\frac{2M-Q^2}{r^2}$ and $\hat{A}=1-\frac{2Mr-Q^2}{r^2+a^2\cos^2\theta}$. More details of the derivation of the Kerr-Newman metric in general relativity can be found in the Appendix A.\\
From eqs.\eqref{general new l} to \eqref{general new m}, we obtain that:
\begin{equation} \label{general ln+nl}
l^{\mu}n^{\nu}+l^{\nu}n^{\mu}=
\left[ \begin{array}{cccc}
		0 & 1 & 0 & 0 \\
		1 & \hat{A} & 0 & 0 \\
		0 & 0 & 0 & 0 \\
		0 & 0 & 0 & 0
\end{array}	\right],
\end{equation}
and 
\begin{equation} \label{general mm+mm}
	m^{\mu}\bar{m}^{\nu}+m^{\nu}\bar{m}^{\mu}=
\left[ \begin{array}{cccc}
		\frac{1}{\widehat{(A^{-1}C)}}\frac{\psi ^2W'^2}{f_1^2}	& \frac{1}{\widehat{(A^{-1}C)}}\frac{\psi ^2W'P'}{f_1^2} & 0 &  \frac{1}{\widehat{(A^{-1}C)}}\frac{\psi W'}{f_1f_2}\\
		\frac{1}{\widehat{(A^{-1}C)}}\frac{\psi ^2W'P'}{f_1^2}	& \frac{1}{\widehat{(A^{-1}C)}}\frac{\psi ^2P'^2}{f_1^2} & 0 & \frac{1}{\widehat{(A^{-1}C)}}\frac{\psi P'}{f_1f_2} \\
		0	& 0 & \frac{1}{\widehat{(A^{-1}C)}f_1^2} & 0 \\
		\frac{1}{\widehat{(A^{-1}C)}}\frac{\psi W'}{f_1f_2}	& \frac{1}{\widehat{(A^{-1}C)}}\frac{\psi P'}{f_1f_2} & 0 & \frac{1}{\widehat{(A^{-1}C)}f_2^2}
	\end{array}	\right],
\end{equation}
Combining \eqref{g up l n m}, \eqref{general ln+nl} and \eqref{general mm+mm}
\begin{equation} \label{general rot metric up}
		g^{\mu \nu}=
\left[ \begin{array}{cccc}
	\frac{1}{\widehat{(A^{-1}C)}}\frac{(\psi W')^2}{f^2_1}	& -1+\frac{1}{\widehat{(A^{-1}C)}}\frac{\psi ^2P'W'}{f^2_1} & 0 & \frac{1}{\widehat{(A^{-1}C)}}\frac{\psi W'}{f_1f_2} \\
	-1+\frac{1}{\widehat{(A^{-1}C)}}\frac{\psi ^2P'W'}{f^2_1}	& - \hat{A}+\frac{1}{\widehat{(A^{-1}C)}}\frac{(\psi P')^2}{f^2_1} & 0 & \frac{1}{\widehat{(A^{-1}C)}}\frac{\psi P'}{f_1f_2} \\
	0	& 0 & \frac{1}{\widehat{(A^{-1}C)}f^2_1} & 0 \\
	\frac{1}{\widehat{(A^{-1}C)}}\frac{\psi W'}{f_1f_2}	& \frac{1}{\widehat{(A^{-1}C)}}\frac{\psi P'}{f_1f_2} & 0 & \frac{1}{\widehat{(A^{-1}C)}}\frac{1}{f^2_2}
	\end{array}	\right].
\end{equation}
Since the metric should be written in terms of line elements, we need to find the inverse of the matrix \eqref{general rot metric up}, which is given by:
\begin{equation} \label{general metric under u,r,theta,phi}
	g_{\mu \nu}=
\left[\begin{array}{cccc}
	-\hat{A}	& -1 & 0 & \hat{A}aW'\frac{f_2}{f_1}+\psi P'\frac{f_2}{f_1}	 \\
	-1	& 0 & 0 & \psi W'\frac{f_2}{f_1} \\
	0	& 0 & \widehat{(A^{-1}C)}f_1^2 & 0 \\
	\hat{A}\psi W'\frac{f_2}{f_1}+\psi P'\frac{f_2}{f_1}	& \psi W'\frac{f_2}{f_1} & 0 & -\hat{A}\psi ^2W'^2\frac{f_2^2}{f_1^2}+\widehat{(A^{-1}C)}f_2^2-2\psi^2P'W'\frac{f_2^2}{f_1^2}
	\end{array}\right].
\end{equation}
In order to cancel the (0,1),(1,0),(1,3),(3,1) components of matrix \eqref{general metric under u,r,theta,phi}, we need to introduce the following transformation:
 \begin{equation} \label{transfer back general}
      \left[\begin{array}{c}
         dt \\
         dr \\
         d\theta\\
         d\varphi
    \end{array}
    \right]=\left[ \begin{array}{cccc}
        1 & -M & 0 & 0\\
        0 & 1 & 0 & 0\\
        0 & 0 & 1 & 0\\
        0 & -N & 0 & 1
    \end{array}
    \right]
    \left[ \begin{array}{c}
        du \\
        dr \\
       d\theta \\
       d\varphi
    \end{array}\right],
 \end{equation}
where
\begin{equation} \label{M}
    M=\frac{-\psi ^2P'W'+\widehat{(A^{-1}C)}f_1^2}{\hat{A}\widehat{(A^{-1}C)}+\psi^2P'^2},
\end{equation}
\begin{equation} \label{N}
    N=\frac{\psi P'}{(\hat{A}\widehat{(A^{-1}C)}+\psi^2P'^2)\left(-\frac{f_2}{f_1}\right)}.
\end{equation}
Applying transformation \eqref{transfer back general}, we can obtain
\begin{equation} \label{last general metric(without dilaton)}
	g_{\mu \nu}=
\left[\begin{array}{cccc}
-\hat{A}	& 0 & 0 & \hat{A} \psi W'\frac{f_2}{f_1}+\psi P'\frac{f_2}{f_1} \\
0	& \frac{\widehat{(A^{-1}C)}f_1^2}{\psi^2P'^2+\hat{A} (\widehat{A^{-1}C})} & 0 & 0 \\
0	& 0 & \widehat{(A^{-1}C)}f_1^2 & 0 \\
\hat{A} \alpha W'\frac{f_2}{f_1}+\psi P'\frac{f_2}{f_1}	& 0 & 0 & -\hat{A}\psi^2W'^2\frac{f_2^2}{f_1^2}+\widehat{(A^{-1}C)}f_2^2-2\psi^2P'W'\frac{f_2^2}{f_1^2}
\end{array}\right].
\end{equation}
Finally, the metric in term of line elements is given by:
\begin{equation} \label{last general metric}
\begin{split}
    ds^2=&-\hat{A}dt^2+\frac{\widehat{(A^{-1}C)}f_1^2}{\psi^2P'^2+\hat{C}}dr^2+ \widehat{(A^{-1}C)}f_1^2 d\theta^2+2\left(\hat{A}\alpha W'\frac{f_2}{f_1}+\psi P'\frac{f_2}{f_1}\right)dtd\varphi \\
    &+\left(-\hat{A}\psi^2W'^2\frac{f_2^2}{f_1^2}+\widehat{(A^{-1}C)}f_2^2-2\psi^2P'W'\frac{f_2^2}{f_1^2}\right)d\varphi^2,
\end{split}
\end{equation}
where $\psi$ has been introduced as a rotation parameter in \eqref{psi}.

\subsection{Kerr-like metric in DFT}
In this section, we first apply the metric derivation method from the previous section to DFT metric. Then we give the  complexification of $A$, $A^{-1}C$ and $C$.\\ 
From \eqref{eq:solution}, we can get
\begin{equation} \label{A(r), C(r)}
	A(r)=\left(\frac{r-c_+}{r-c_-}\right)^{\frac{a}{c_+-c_-}} \qquad C(r)=(r-c_+)(r-c_-),
\end{equation}
For the Kerr-like metric in DFT, we set:
\begin{equation} \label{r u dft}
    r'=r+i\psi \cos{\theta} \qquad u'=u-i\psi \cos{\theta}.
\end{equation}
Taking the derivative of \eqref{r u dft}, we can get
\begin{equation} \label{dr dft}
    dr'=dr-i\psi \sin{\theta} \qquad du'=u+i\psi \sin{\theta},
\end{equation}
so we have 
\begin{equation} \label{f1 f2 P' W' in dft}
    f_1=1, \quad f_2=\sin \theta \quad \text{and} \quad P'=-\sin \theta, \quad W'=\sin \theta. 
\end{equation}
Insert \eqref{f1 f2 P' W' in dft} into \eqref{last general metric(without dilaton)}, then we have
\begin{equation} \label{the general Kerr-like metric in DFT without dilaton}
g_{\mu \nu}=
\left[\begin{array}{cccc}
-\hat{A}	& 0 & 0 & \psi \sin^2{\theta} (\hat{A}-1)	 \\
0	& \frac{\widehat{A^{-1}C}}{\hat{A}\widehat{(A^{-1}C)}+\psi^2 \sin^2\theta} & 0 & 0 \\
0	& 0 & \widehat{(A^{-1}C)} & 0 \\
\psi \sin^2{\theta} (\hat{A}-1)	& 0 & 0 & -\hat{A}\psi^2 \sin^4{\theta}+\widehat{(A^{-1}C)} \sin^2{\theta}+2\psi^2 \sin^4{\theta}
\end{array}
\right].
\end{equation}
where $\hat{A}$, $\widehat{(A^{-1}C)}$ will be given by \eqref{hat A} and \eqref{hat AC}.\\
Moreover, considering \eqref{M} and \eqref{N}, we can obtain the corresponding $M$ and $N$ in 'Kerr-like' metric directly,
\begin{equation} \label{M DFT}
    M=\frac{\widehat{(A^{-1}C)}+\psi^2\sin^2\theta}{\hat{A}\widehat{(A^{-1}C)}+\psi^2\sin^2\theta},
\end{equation}
\begin{equation} \label{N DFT}
    N=\frac{\psi}{\hat{A}\widehat{(A^{-1}C)}+\psi^2\sin^2\theta}.
\end{equation}
In order to find the metric without dilaton, we also need to do complexification to get $\hat{A}$  and $\widehat{A^{-1}C}$.
For easier calculation, we first consider the special condition with $c_-=0$:
\begin{equation}
	A(r)|_{c_-=0}=\left(1-\frac{\sqrt{a^2+b^2}}{r}\right)^\frac{a}{\sqrt{a^+b^2}}.
\end{equation}
The complexification of $A(r)|_{c_-=0}$ is:
\begin{equation}
	\hat{A}(r) |_{c_-=0}=\left(\frac{r^2+\psi^2 \cos^2{\theta}-r\sqrt{a^2+b^2}}{r^2+\psi^2 \cos^2{\theta}}\right)^\frac{a}{\sqrt{a^+b^2}}.
\end{equation}
We replace $r$ by $r-c_-$, which also satisfy \eqref{dr dft}, so we can get:
\begin{equation} \label{hat A}
	\hat{A}(r)=\left(\frac{(r-c_-)^2+\psi^2 \cos^2{\theta}-(r-c_-)\sqrt{a^2+b^2}}{(r-c_-)^2+\psi^2 \cos^2{\theta}}\right)^\frac{a}{\sqrt{a^2+b^2}},
\end{equation}
Similarly, we can get the complexification of $A^{-1}C$ and $C$:
\begin{equation} \label{hat AC}
	\widehat{(A^{-1}C)}=((r-c_-)^2+\psi^2 \cos^2{\theta}-(r-c_-)\sqrt{a^2+b^2})^{1-\frac{a}{\sqrt{a^2+b^2}}}((r-c_-)^2+\psi^2 \cos^2{\theta})^{\frac{a}{\sqrt{a^2+b^2}}},
\end{equation}
\begin{equation} \label{hat C}
	\hat{C}=(r-c_-)^2+\psi^2 \cos^2{\theta}-(r-c_-)\sqrt{a^2+b^2}.
\end{equation}
When we consider the complete Schwarzschild metric in string frame \cite{KPS2017}
\begin{equation}
   \begin{split}
        ds^2&=-(\gamma_+(\frac{r-c_+}{r-c_-})^{\frac{a+b}{\sqrt{a^2+b^2}}}+\gamma_-(\frac{r-c_+}{r-c_-})^{\frac{a-b}{\sqrt{a^2+b^2}}})dt^2\\
        &+(\gamma_+(\frac{r-c_+}{r-c_-})
    ^{\frac{-a+b}{\sqrt{a^2+b^2}}}+\gamma_-(\frac{r-c_+}{r-c_-})^{-\frac{a+b}{\sqrt{a^2+b^2}}})dr^2\\
    &+(\gamma_+(r-c_+)^{1+\frac{-a+b}{\sqrt{a^2+b^2}}}(r-c_-)^{1+\frac{a-b}{a^2+b^2}}+\gamma_-(r-c_-)^{1-\frac{a+b}{\sqrt{a^2+b^2}}}(r-c_+)^{1+\frac{a+b}{\sqrt{a^2+b^2}}}d\theta^2\\
    &+(\gamma_+(r-c_+)^{1+\frac{-a+b}{\sqrt{a^2+b^2}}}(r-c_-)^{1+\frac{a-b}{a^2+b^2}}+\gamma_-(r-c_-)^{1-\frac{a+b}{\sqrt{a^2+b^2}}}(r-c_+)^{1+\frac{a+b}{\sqrt{a^2+b^2}}}\sin^2\theta d\varphi^2,
   \end{split}
\end{equation}
we cannot directly give a complete metric, that is, we cannot give the specific form of the dilaton corresponding to this metric. Now we denote the corresponding dilaton as $e^{2\hat{\phi}}$.\\
Now, based on \cite{EA and JC 2001}, we can write the general Kerr-like metric in DFT in string frame is
\begin{equation} \label{line element of the general Kerr-like metric in DFT}
\begin{split}
    ds^2 &=e^{2\hat{\phi}}(-\hat{A}dt^2+\left(\frac{\widehat{(A^{-1}C)}}{{\hat{A}}\widehat{(A^{-1}C)}+\psi^2 \sin^2\theta}\right)dr^2+\widehat{(A^{-1}C)}d\theta^2+(-\hat{A}\psi^2 \sin^4{\theta}+\widehat{(A^{-1}C)} \sin^2{\theta} \\ 
    &+2\psi^2 \sin^4{\theta})d\varphi^2+\psi \sin^2{\theta} (\hat{A}-1)2dt d\varphi).
\end{split}
\end{equation}

\subsection{Discussion about dilaton and $H$-flux}
In this subsection, we consider the conditions which $H$-flux and dilaton should satisfy. For Kerr-like metric, we can set the general form of $B_{(2)}$:
\begin{equation} \label{general B_2}
    B_{(2)}=Udt\wedge dr+Vdt \wedge d\theta+Wdt \wedge d\varphi+Xdr \wedge d\theta+Ydr \wedge d\varphi+Zd\theta \wedge d\varphi
\end{equation}
where $U$,$V$,$W$,$X$,$Y$,$Z$ are all the functions of $r$ and $\theta$. The explicit expressions for these functions should be determined from string theory \cite{KPS2017, K.S. STELLE 1997}, but we will not discuss it in this paper. From \eqref{general B_2} we give $H_{(3)}$:
\begin{equation} \label{general H_3}
    H_{(3)}=\left( \frac{\partial U}{\partial \theta}-\frac{\partial V}{\partial r}\right)dt \wedge dr \wedge d\theta-\frac{\partial W}{\partial r}dt \wedge dr \wedge d\varphi-\frac{\partial W}{\partial \theta} dt \wedge d\theta \wedge d\varphi+\left(-\frac{\partial Y}{\partial \theta}+\frac{\partial Z}{\partial r}\right)dr \wedge d\theta \wedge d\varphi,
\end{equation}
For subsequent calculations, we assume that the dilaton $\hat{\phi}$ is a function of $r$ and $\theta$ due to Kerr-like metric. Now we calculate \eqref{eq: generalsed Einstein equation 2}
\begin{equation} \label{calculate d hodge}
    \begin{split}
        d \star (e^{2\hat{\phi}}H_{(3)})
        &=d(e^{2\hat{\phi}}\left(\frac{\partial U}{\partial \theta}-\frac{\partial V}{\partial r}\right)(-\sqrt{|g|}g^{\varphi \varphi})d\varphi-e^{2\hat{\phi}}\frac{\partial W}{\partial r}\sqrt{|g|}g^{\theta \theta}d\theta\\
        &-e^{2\hat{\phi}}\frac{\partial W}{\partial \theta}\sqrt{|g|}g^{rr}dr+e^{2\hat{\phi}}\left(-\frac{\partial Y}{\partial \theta}+\frac{\partial Z}{\partial r}\right)\sqrt{|g|}g^{tt}dt)\\
        &=-\frac{\partial}{\partial r}\left(e^{2\hat{\phi}}\left(-\frac{\partial Y}{\partial \theta}+\frac{\partial Z}{\partial r}\right)\sqrt{|g|}g^{tt}\right)dt\wedge dr-\frac{\partial}{\partial \theta}\left(e^{2\hat{\phi}}\left(-\frac{\partial Y}{\partial \theta}+\frac{\partial Z}{\partial r}\right)\sqrt{|g|}g^{tt}\right) dt \wedge d\theta\\
        &+\left(-\frac{\partial}{\partial \theta}\left(-e^{2\hat{\phi}}\frac{\partial W}{\partial \theta}\sqrt{|g|}g^{rr}\right)+\frac{\partial}{\partial r}\left(-e^{2\hat{\phi}}\frac{\partial W}{\partial r}\sqrt{|g|}g^{\theta \theta}\right)\right)dr\wedge d\theta\\
        &+\frac{\partial}{\partial r}\left(e^{2\hat{\phi}}\left(\frac{\partial U}{\partial \theta}-\frac{\partial V}{\partial r}\right)(-\sqrt{|g|}g^{\varphi \varphi})\right)dr\wedge d\varphi+\frac{\partial}{\partial \theta}\left(e^{2\hat{\phi}}\left(\frac{\partial U}{\partial \theta}-\frac{\partial V}{\partial r}\right)(-\sqrt{|g|}g^{\varphi \varphi})\right)d\theta \wedge d\varphi.
    \end{split}
\end{equation}
We calculate some components in metric:
 \begin{equation} \label{a size of components of g}
     \begin{split}
         g^{tt}&=e^{-2\hat{\phi}}\frac{-\hat{C}+\psi^2(-2+\hat{A})\sin^2\theta}{\hat{A}\hat{C}+\psi^2\sin^2\theta}\\
         g^{rr}&=e^{-2\hat{\phi}}\frac{\hat{A}(\hat{C}+\psi^2\sin^2\theta)}{\hat{C}}\\
         g^{\theta \theta}&=e^{-2\hat{\phi}}\frac{\hat{A}}{\hat{C}}\\
         g^{\varphi \varphi}&=e^{-2\hat{\phi}}\frac{\hat{A}\csc^4\theta}{\psi^2+\hat{A}\csc^2\theta\hat{C}}\\
         \sqrt{|g|}&=e^{4\hat{\phi}}\frac{\hat{C}\sin^2\theta}{\hat{A}}\sqrt{\frac{\hat{A}\hat{C}+\psi^2\sin^2\theta}{\hat{C}+\psi^2\sin^2\theta}}
     \end{split}
 \end{equation}
and some terms in $d \star (e^{2\hat{\phi}}H_{(3)})=0$:
\begin{equation} \label{some terms in H_3}
    \begin{split}
         e^{2\hat{\phi}}\left(-\frac{\partial Y}{\partial \theta}+\frac{\partial Z}{\partial r}\right)\sqrt{|g|}g^{tt}&= \left(-\frac{\partial Y}{\partial \theta}+\frac{\partial Z}{\partial r}\right)P_1\\
         e^{2\hat{\phi}}\frac{\partial W}{\partial \theta}\sqrt{|g|}g^{rr}&=-\frac{\partial W}{\partial \theta}P_2 \\
        e^{2\hat{\phi}}\frac{\partial W}{\partial r}\sqrt{|g|}g^{\theta \theta}&=\frac{\partial W}{\partial r}P_3 \\
        e^{2\hat{\phi}}\left(\frac{\partial U}{\partial \theta}-\frac{\partial V}{\partial r}\right)(-\sqrt{|g|}g^{\varphi \varphi})&=\left(\frac{\partial U}{\partial \theta}-\frac{\partial V}{\partial r}\right)P_4
    \end{split}
\end{equation}
where 
\begin{equation} \label{donete g'^tt,g^'rr,g'^thetatheta,g'^varphivarphi}
    \begin{split}
        P_1&=\left(\frac{-\hat{C}+\psi^2(-2+\hat{A})\sin^2\theta}{\hat{A}\hat{C}+\psi^2\sin^2\theta}\right)\left(\frac{\hat{C}\sin^2\theta}{\hat{A}}\right)\sqrt{\frac{\hat{A}\hat{C}+\psi^2\sin^2\theta}{\hat{C}+\psi^2\sin^2\theta}} \\
        P_2&=\left(\frac{\hat{A}(\hat{C}+\psi^2\sin^2\theta)}{\hat{C}}\right)\left(\frac{\hat{C}\sin^2\theta}{\hat{A}}\right)\sqrt{\frac{\hat{A}\hat{C}+\psi^2\sin^2\theta}{\hat{C}+\psi^2\sin^2\theta}} \\
       P_3&=\left(\frac{\hat{A}}{\hat{C}}\right)\left(\frac{\hat{C}\sin^2\theta}{\hat{A}}\right)\sqrt{\frac{\hat{A}\hat{C}+\psi^2\sin^2\theta}{\hat{C}+\psi^2\sin^2\theta}} \\
        P_4&=\left(\frac{\hat{A}\csc^4\theta}{\psi^2+\hat{A}\csc^2\theta\hat{C}} \right) \left(\frac{\hat{C}\sin^2\theta}{\hat{A}}\right)\sqrt{\frac{\hat{A}\hat{C}+\psi^2\sin^2\theta}{\hat{C}+\psi^2\sin^2\theta}}
    \end{split}
\end{equation}
Then we need to solve a series of equations from \eqref{general H_3}
\begin{equation} \label{equations form H_3}
    \begin{split}
        \left(-\frac{\partial Y}{\partial \theta}+\frac{\partial Z}{\partial r}\right)P_1&=c_1\\
        \left(\frac{\partial U}{\partial \theta}-\frac{\partial V}{\partial r}\right)P_4&=c_2\\
        \frac{\partial}{\partial \theta}\left(\frac{\partial W}{\partial \theta}\right)P_2&=\frac{\partial}{\partial r}\left(\frac{\partial W}{\partial r}\right)P_3
    \end{split}
\end{equation}
where $c_1$,$c_2$ are the constants. We can also calculate the components of $H_{\mu \rho \sigma}H_{\nu}^{\ \rho \sigma}$:
\begin{equation} \label{components of HH 1}
    \begin{split}
       H_{t\rho \sigma}H_t^{\ \rho \sigma}&=2\left(\left(\frac{\partial U}{\partial \theta}-\frac{\partial V}{\partial r}\right)^2g^{\theta\theta}g^{rr}+\left(\frac{\partial W}{\partial r}\right)^2g^{rr}g^{\varphi\varphi}+\left(\frac{\partial W}{\partial \theta}\right)^2g^{\theta\theta}g^{\varphi\varphi}\right)\\
       H_{t\rho \sigma}H_r^{\ \rho \sigma}&=-2\frac{\partial W}{\partial \theta}\left(-\frac{\partial Y}{\partial \theta}+\frac{\partial Z}{\partial r}\right)g^{\theta\theta}g^{\varphi\varphi}\\
       H_{t \rho \sigma}H_{\theta}^{\ \rho \sigma}&=-2\frac{\partial W}{\partial \theta}\left(-\frac{\partial Y}{\partial \theta}+\frac{\partial Z}{\partial r}\right)g^{rr}g^{\varphi\varphi}\\
       H_{t \rho \sigma}H_{\varphi}^{\ \rho \sigma}&=2\left(\frac{\partial U}{\partial \theta}-\frac{\partial V}{\partial r}\right)\left(-\frac{\partial Y}{\partial \theta}+\frac{\partial Z}{\partial r}\right)g^{rr}g^{\varphi\varphi}\\
       H_{r\rho \sigma}H_r^{\ \rho \sigma}&=2\left(\left(\frac{\partial U}{\partial \theta}-\frac{\partial V}{\partial r}\right)^2g^{tt}g^{\theta\theta}+\left(\frac{\partial W}{\partial \theta}\right)^2g^{tt}g^{\varphi\varphi}+\left(-\frac{\partial Y}{\partial \theta}+\frac{\partial Z}{\partial r}\right)^2g^{\theta\theta}g^{\ \varphi\varphi}\right)\\
       H_{r\rho \sigma}H_{\theta}^{\ \rho \sigma}&=2\frac{\partial W}{\partial \theta}\frac{\partial W}{\partial r}g^{tt}g^{\varphi\varphi}\\
       H_{r\rho \sigma}H_{\varphi}^{\ \rho \sigma}&=2(\left(\frac{\partial U}{\partial \theta}-\frac{\partial V}{\partial r}\right)\left(\frac{\partial W}{\partial r}\right)g^{tt}g^{\theta\theta}\\
       H_{\theta \rho \sigma}H_{\theta}^{\ \rho \sigma}&=2\left(\left(\frac{\partial U}{\partial \theta}-\frac{\partial V}{\partial r}\right)^2g^{tt}g^{rr}+\left(\frac{\partial W}{\partial \theta}\right)^2g^{tt}g^{\varphi\varphi}+\left(-\frac{\partial Y}{\partial \theta}+\frac{\partial Z}{\partial r}\right)^2g^{rr}g^{\varphi\varphi}\right)\\
       H_{\theta \rho \sigma}H_{\varphi}^{\ \rho \sigma}&=-2\left(\frac{\partial U}{\partial \theta}-\frac{\partial V}{\partial r}\right)\frac{\partial W}{\partial r}g^{tt}g^{rr}\\
       H_{\varphi \rho \sigma}H_{\varphi}^{\ \rho \sigma}&= 2\left(\left(\frac{\partial W}{\partial r}\right)^2g^{tt}g^{rr}+\left(\frac{\partial W}{\partial \theta}\right)^2g^{tt}g^{\theta\theta}+\left(-\frac{\partial Y}{\partial \theta}+\frac{\partial Z}{\partial r}\right)^2g^{rr}g^{\theta\theta}\right)
    \end{split}
\end{equation}
Now we deal with $\nabla_{\mu}(\partial_{\nu}\phi)$:
\begin{equation} \label{nabla mu partial nu phi}
\left[
    \begin{array}{cccc}
       -(\Gamma^r_{tt}\partial_r\phi+\Gamma^{\theta}_{tt}\partial_{\theta}\phi) & 0 & 0 & -(\Gamma^r_{t\varphi}\partial_r\phi+\Gamma^{\theta}_{t\varphi}\partial_{\theta}\phi) \\
       0 & \partial_r\partial_r\phi-\Gamma^r_{rr}\partial_r\phi-\Gamma^{\theta}_{rr}\partial_{\theta}\phi & \partial_{\theta}\partial_r\phi-\Gamma^r_{r\theta}\partial_r\phi-\Gamma^{\theta}_{r\theta}\partial_{\theta}\phi & 0\\
       0 & \partial_r\partial_{\theta}\phi-\Gamma^r_{r\theta}\partial_r\phi-\Gamma^{\theta}_{r\theta}\partial_{\theta}\phi & \partial_r\partial_r\phi-\Gamma^r_{\theta\theta}\partial_r\phi-\Gamma^{\theta}_{\theta\theta}\partial_{\theta}\phi & 0\\
    -(\Gamma^r_{t\varphi}\partial_r\phi+\Gamma^{\theta}_{t\varphi}\partial_{\theta}\phi) & 0 & 0 & -(\Gamma^r_{\varphi\varphi}\partial_r\phi+\Gamma^{\theta}_{\varphi\varphi}\partial_{\theta}\phi)
    \end{array}\right],
\end{equation}
where $\Gamma$ is Christoffel symbol. We will list the non-vanishing components of the Christoffel symbol in Appendix B. Regarding the curvature tensor, due to its complexity, we will not list explicit results, but we can piont out that $R_{tt}$, $R_{rr}$, $R_{\theta\theta}$, $R_{\varphi\varphi}$, $R_{t\varphi}=R_{\varphi t}$ and $R_{r\theta}=R_{\theta r}$ are the non-vanishing components of Ricci curvature. Based on \eqref{components of HH 1} and \eqref{nabla mu partial nu phi},we have:
\begin{equation} \label{limit H_(3)}
    \begin{split}
        \frac{\partial W}{\partial \theta}\left(-\frac{\partial Y}{\partial \theta}+\frac{\partial Z}{\partial r}\right)&=0\\
        \left(\frac{\partial U}{\partial \theta}-\frac{\partial V}{\partial r}\right)\frac{\partial W}{\partial r}&=0
    \end{split}
\end{equation}
We have $\frac{\partial W}{\partial \theta}=0$ or $-\frac{\partial Y}{\partial \theta}+\frac{\partial Z}{\partial r}=0$ and $\frac{\partial U}{\partial \theta}-\frac{\partial V}{\partial r}=0$ or $\frac{\partial W}{\partial r}=0$. 
Since this metric can degenerate to the solution of spherically symmetric ansatz (under the condition of $\psi=0$), we have:
\begin{equation}
    H_{(3)}|_{\psi=0}=h\sin\theta dt\wedge d\theta \wedge d\varphi
\end{equation}

\section{Black hole thermodynamics and Hawking radiation}
In this section, we elaborate on the first law of thermodynamics of black holes in DFT based on subsection 2.3. Then we give the explicit form of the first law of thermodynamics of the metric we have obtained in subsection 3.2. Afterwards, we also calculate the Hawking radiation for the $4D$ DFT Kerr-like metric in detail.

\subsection{The first law of the black-hole thermodynamics of the DFT Kerr-like metric}
Let us consider the thermodynamics of black holes of the $4D$ DFT Kerr-like metric, which is given by \eqref{last general metric}.\\
\textbf{Part 1. Entropy term:}\\
To find the entropy term, we need to obtain the event horizon, which corresponds to "Cauchy surface" mentioned in subsection 2.3. The first step is to find its radius, namely, the pole of \eqref{M} and \eqref{N}:
\begin{equation} \label{radius}
	\hat{A}\widehat{(A^{-1}C)}+\psi^2\sin^2\theta=0.
\end{equation}
By substituting the two terms $\hat{A}$ \eqref{hat A} and $\widehat{A^{-1}C}$ \eqref{hat AC} into \eqref{radius}, we can obtain that
\begin{equation}
	\left((r-c_-)^2+\psi^2 \cos^2{\theta} -(r-c_-)\sqrt{a^2+b^2}\right)+\psi^2\sin^2\theta=0.
\end{equation}
Similar to the Kerr metric in general relativity, this equation has a maximum solution, denoted as $r=r_+$, which can lead to an event horizon, on which thermodynamics exists. Now, we give the event horizon in the following equation:\\
\begin{equation}
	\mathcal{A}=\int_{r_+}\sqrt{g_{\theta \theta} g_{\varphi \varphi}}d\theta d\varphi=\int_{r_+} e^{2\hat{\phi}}\sqrt{{\widehat{(A^{-1}C)}} (-\hat{A}\psi^2 \sin^4{\theta}+\widehat{(A^{-1}C)} \sin^2{\theta}+2\psi^2 \sin^4{\theta})}
\end{equation}
From Bekenstein-Hawking entropy formula, we can get the entropy
\begin{equation} \label{entropy 2}
	S=\frac{1}{4G_{DFT}}\mathcal{A}
\end{equation}
\textbf{Part 2. $\delta M'$ and $\slashed{\delta}W$:}\\
As discussed earlier, we can give the remaining two items $\delta M'$ and $\slashed{\delta}W$. Before that, we need to get $\Omega^I$ and $J_I$. Similar to general relativity \cite{R.M.Wald 1984}, we can directly have
\begin{equation} \label{J I}
	J_I \equiv (\frac{1}{16 \pi G_{DFT}})\int_{t=t_0, r=r_+}e^{2\hat{\phi}}  \epsilon_{abcd}\nabla^c \xi^d.
\end{equation} 
Moverover, as a Killing field, $\xi^d$ is defined as
\begin{equation} \label{xi d} 
    \xi^d=\frac{\partial}{\partial t}+\Omega^I \frac{\partial}{\partial \varphi^I}
\end{equation}
To give $\Omega^I$, we need to first define $\Omega$. By taking the limit $r \rightarrow r_+$, $\Omega$ approaches $\Omega^I$, namely,
\begin{equation} \label{Omega}
\begin{split}
   \Omega^I&= \lim_{r \rightarrow r_+} \Omega =\lim_{r \rightarrow r_+} \frac{d\varphi}{dt}= \lim_{r \rightarrow r_+} \left(-\frac{g_{t\varphi}}{g_{\varphi \varphi}} \right) \\
   &= \lim_{r \rightarrow r_+} \left(-\frac{\psi \sin^2\theta (\hat{A}-1)}{-\hat{A}\psi^2\sin^4\theta+\widehat{(A^{-1}C)}\sin^2\theta+2\psi^2\sin^4\theta} \right).    
\end{split}
\end{equation}
and based on \eqref{slashed delta W}, we can define $\slashed{\delta}W$, which is $O(n, n)$ invariant thermodynamic work contribution corresponding to metric \eqref{the general Kerr-like metric in DFT without dilaton}
\begin{equation} \label{new W}
\begin{split}
      \slashed{\delta}W=&-\frac{1}{16\pi G_{DFT}} \int_{t=t_0, r=r_+}d^2x[\tilde{\lambda}_{\rho}\delta(e^{2\hat{\phi}}  \mathcal{H}^{\mu \nu \rho}+\tilde{\Lambda}^M \delta(e^{2\hat{\phi}} H_{MN} \mathcal{F}^{\mu \nu N})]\epsilon_{\mu \nu}\\
      &-\Omega^I \left(\frac{1}{16 \pi G_{DFT}}\right) \delta \int_{t=t_0 ,r=r_+}e^{2\hat{\phi}}  \epsilon_{abcd}\nabla^c \xi^d.
\end{split}
\end{equation}
Then we can get $\delta M'$, which is
\begin{equation} \label{delta M'}
\begin{split}
    \delta M'=&\delta M-\Omega^I\delta J_I =\delta M-\Omega^I \left(\frac{1}{16 \pi G_{DFT}}\right) \delta \int_{t=t_0, r=r_+}\epsilon_{abcd}\nabla^c \xi^d \\
    &=\frac{1}{16 \pi G_{DFT}} \delta\int_{t=t_0, r=r_+} d^2x e^{2\hat{\phi}}(-2g^{\rho[\mu}\nabla_{\rho}(\xi')^{\nu ]}-\widetilde{\lambda}_pH^{\mu \nu \rho}-\widetilde{\Lambda}^M H_{MN}\mathcal{F}^{\mu \nu N})\epsilon_{\mu\nu}\\
    & -\Omega^I \left(\frac{1}{16 \pi G_{DFT}}\right) \delta \int_{t=t_0, r=r_+} e^{2\hat{\phi}}  \epsilon_{abcd}\nabla^c \xi^d.
\end{split}
\end{equation}
Moreover, based on discussion about the "Komar term" \cite{ASA and CDAB 2017}, we can rewrite \eqref{delta M'} as
\begin{equation} \label{delta M'2}
\begin{split}
    \delta M'&= \frac{\kappa}{16 \pi G_{DFT}} \delta \int_{t=t_0, r=r_+} e^{2\hat{\phi}}\sqrt{g_{\theta \theta}g_{\varphi \varphi}}\\
    & -\frac{1}{16\pi G_{DFT}} \int_{t=t_0, r=r_+}d^2x[\tilde{\lambda}_{\rho}\delta(e^{2\hat{\phi}} \mathcal{H}^{\mu \nu \rho}+\tilde{\Lambda}^M \delta(e^{2\hat{\phi}} \mathcal{F}^{\mu \nu N})]\epsilon_{\mu \nu}\\
      &-\Omega^I \left(\frac{1}{16 \pi G_{DFT}}\right) \delta \int_{t=t_0 ,r=r_+}e^{2\hat{\phi}}  \epsilon_{abcd}\nabla^c \xi^d,\\
    &=\frac{\kappa}{2\pi}\delta S
    -\frac{1}{16\pi G_{DFT}} \int_{t=t_0, r=r_+}d^2x[\tilde{\lambda}_{\rho}\delta(e^{2\hat{\phi}} \mathcal{H}^{\mu \nu \rho}+\tilde{\Lambda}^M \delta(e^{2\hat{\phi}}H_{MN} \mathcal{F}^{\mu \nu N})]\epsilon_{\mu \nu}\\
      &-\Omega^I \left(\frac{1}{16 \pi G_{DFT}}\right) \delta \int_{t=t_0 ,r=r_+}e^{2\hat{\phi}}  \epsilon_{abcd}\nabla^c \xi^d,
\end{split}
\end{equation}
Combining \eqref{new W} and \eqref{delta M'2}, the first law of black-hole thermodynamics \eqref{first law 2} can be obtained.

\subsection{Hawking radiation as tunneling}
In this subsection, we study the Hawking radiation of the Kerr-metric we have obtained in \eqref{last general metric}. In line with the analysis of \cite{Parikh&Wilczek:2000, Johnson&March-Russell:2020}, we study the semiclassical tunnelling of particles through the horizon of black holes in this section. Further discussion can be seen in \cite{Kraus&Wilczek:1995, Parikh:2004, Parikh:2003, Vanzo&Acquaviva&Criscienzo:2011}. On the one hand, the rate of emission $\Gamma$ will have exponential part given by \cite{Johnson&March-Russell:2020}
\begin{equation} \label{GammaS}
\Gamma \sim \exp(-2 Im S),
\end{equation}
where $S$ is the tunnelling action of particles and $ImS$ is the imaginary part of the tunnelling action $S$. On the other hand, according to the Planck radiation law, the emitted rate $\Gamma$ of particles with frequency $\omega$ can be written as \cite{Johnson&March-Russell:2020}
\begin{equation} \label{Gammaomega}
\Gamma \sim \exp(-\omega/T_{BH}).
\end{equation}
Therefore, combining eqs.(\ref{GammaS}) and (\ref{Gammaomega}), the temperature
at which the black hole radiates can be read off \cite{Johnson&March-Russell:2020}:
\begin{equation} \label{TomegaS}
T_{BH}=\frac{\omega}{2ImS}.
\end{equation}
In this section, we study Hawking radiation process in Painlevé–Gullstrand coordinates. Firstly, we will rewrite the metric \eqref{the general Kerr-like metric in DFT without dilaton} in Painlevé–Gullstrand coordinates. We define 
\begin{equation} \label{ts}
t_s =t - F(r,\rho),
\end{equation}
where $F(r,\rho)$ is a function of $r$ and $\rho$. Then we have
\begin{equation} \label{dt}
dt=dt_s + dF(r,\rho)=dt_s+\frac{\partial F(r,\rho)}{\partial r} dr + \frac{\partial F(r,\rho)}{\partial \rho} d\rho,
\end{equation}
and
\begin{equation} \label{dt2}
\begin{split}
dt^2=dt^2_s + \left(\frac{\partial F(r,\rho)}{\partial r}\right)^2 dr^2 + \left(\frac{\partial F(r,\rho)}{\partial \rho}\right)^2 d\rho^2 \\
+2\frac{\partial F(r,\rho)}{\partial r} dt_s dr + 2\frac{\partial F(r,\rho)}{\partial \rho} dt_s d\rho + 2\frac{\partial F(r,\rho)}{\partial r} \frac{\partial F(r,\rho)}{\partial \rho} dr d\rho. \\
\end{split}
\end{equation}
The metric \eqref{the general Kerr-like metric in DFT without dilaton} can be written as
\begin{equation} \label{metricds2}
ds^2=g_{00}dt^2 + g_{11}dr^2 + g_{22} d\rho^2 + g_{33}d\theta^2 + 2 g_{03}dt d\theta.
\end{equation}
In order to obtain the metric in Painlevé–Gullstrand coordinates, we can set \cite{Jiang&Wu&Cai:2006}
\begin{equation} \label{dthetadt}
d\theta=-\frac{g_{03}}{g_{33}}dt.
\end{equation}
Then consider (\ref{dt}) and \eqref{dt2} and ignore $\rho$ sector for simplicity, we have 
\begin{equation} \label{ds2PG}
ds^2=\left(g_{00}-\frac{g^2_{03}}{g_{33}}\right)dt^2_s+\left[\left(g_{00}-\frac{g^2_{03}}{g_{33}}\right)\left( \frac{\partial F(r,\rho)}{\partial r}\right)^2 +g_{11} \right]dr^2 + 2\left(g_{00}-\frac{g^2_{03}}{g_{33}}\right) \frac{\partial F(r,\rho)}{\partial r} dt_s dr.
\end{equation}
If we set
\begin{equation} \label{dr2eq1}
\left(g_{00}-\frac{g^2_{03}}{g_{33}}\right)\left( \frac{\partial F(r,\rho)}{\partial r}\right)^2 +g_{11}=1,
\end{equation}
then we have
\begin{equation} \label{pFpr}
\frac{\partial F(r,\rho)}{\partial r}=\sqrt{\frac{1-g_{11}}{g_{00}-\frac{g^2_{03}}{g_{33}}}}.
\end{equation}
Then the metric (\ref{ds2PG}) can be written as
\begin{equation} \label{ds2PG2}
ds^2=-\left(\frac{g^2_{03}}{g_{33}} - g_{00}\right)dt^2_s+  2\sqrt{\left(g_{00}-\frac{g^2_{03}}{g_{33}} \right) \left(1-g_{11}\right)} dt_s dr + dr^2.
\end{equation}
In the following, we take $dt^2_s=dt^2$ for convenience, namely,
\begin{equation} \label{ds2PG3}
ds^2=-\left(\frac{g^2_{03}}{g_{33}} - g_{00}\right)dt^2+  2\sqrt{\left(g_{00}-\frac{g^2_{03}}{g_{33}} \right) \left(1-g_{11}\right)} dt dr + dr^2.
\end{equation}

The action for a particle moving freely in a curved background can be written as:
\begin{equation} \label{Scb1}
S=\int p_{\mu} dx^{\mu},
\end{equation}
with
\begin{equation} \label{pmu}
p_{\mu}=mg_{\mu\nu}\frac{dx^{\nu}}{d\sigma},
\end{equation}
where $\sigma$ is an affine parameter along the worldline of the particle, chosen so that $p_{\mu}$ coincides with the physical 4-momentum of the particle. For a massive particle, this requires that $d\sigma = d\tau/m$, with $\tau$ the proper time \cite{Johnson&March-Russell:2020, YL2022}. For simplicity, in this paper we only consider the case of massless scalar field $\Phi$ and ignore the angular directions. Following the spirit of \cite{Johnson&March-Russell:2020}, we will obtain the formula of Hawking temperature for the 'Kerr-like' solution in $D = 4$ double field theory.\\
The radial dynamics of massless particles in 4d spacetime are determined by the equations:
\begin{equation} \label{eom1}
\left(\frac{g^2_{03}}{g_{33}}-g_{00}\right)\dot{t}^2-2\sqrt{\left(g_{00}-\frac{g^2_{03}}{g_{33}} \right) \left(1-g_{11}\right)} \dot{t} \dot{r} - \dot{r}^2=0,
\end{equation}
\begin{equation} \label{eom2}
\left(\frac{g^2_{03}}{g_{33}}-g_{00}\right)\dot{t}-\sqrt{\left(g_{00}-\frac{g^2_{03}}{g_{33}} \right) \left(1-g_{11}\right)} \dot{r} = \omega.
\end{equation}
The second equation is the geodesic equation corresponding
to the time-independence of the metric; in terms of the
momentum defined in eq.(\ref{pmu}), it can be written $p_t=-\omega$, and so $\omega$ has the interpretation of the energy of the particle as measured at infinity \cite{Johnson&March-Russell:2020, YL2022}.

In the following, we shall investigate the tunneling behavior of massless particles from the horizon. For this purpose, let us first evaluate the radial, null geodesics. Since the tunneling processes take place near the event horizon, we may consider a particle tunneling across the event horizon as an ellipsoid shell and hold that the particle should still be an ellipsoid shell during the tunneling process. In other words, the particle does not have motion in the $\rho$-direction \cite{Jiang&Wu&Cai:2006}. Therefore, under these assumptions ($ds^2 = 0 = d\rho$), the radial, null geodesics followed by massless particles are given by the following.

According to \eqref{eom1} and \eqref{eom2}, for an outgoing massless particle, we have
\begin{equation} \label{drdt}
\frac{dr}{dt}=\frac{\dot{r}}{\dot{t}}=-\sqrt{\left(g_{00}-\frac{g^2_{03}}{g_{33}} \right) (1-g_{11})} + \sqrt{-g_{11} \left(g_{00}-\frac{g^2_{03}}{g_{33}} \right)},
\end{equation}
\begin{equation} \label{dott}
\dot{t}= \frac{\omega}{\left(\frac{g^2_{03}}{g_{33}}-g_{00}\right)[\sqrt{-g_{11}(1-g_{11})}+g_{11}]},
\end{equation}
and
\begin{equation} \label{dotr}
\dot{r}= \frac{-\omega}{\sqrt{g_{00}-\frac{g^2_{03}}{g_{33}}}} \frac{\sqrt{-g_{11}} - 2g_{11} \sqrt{1-g_{11}}  }{g_{11}}.
\end{equation}
For the case we are considering, we have
\begin{equation} \label{pr}
p_r=g_{rt}\dot{t} + g_{rr}\dot{r} = \frac{\omega}{\sqrt{g_{00}-\frac{g^2_{03}}{g_{33}}}} \left( \frac{\sqrt{1-g_{11}} }{g_{11} + \sqrt{-g_{11} (1-g_{11})}} - \frac{\sqrt{-g_{11}} - 2g_{11} \sqrt{1-g_{11}}  }{g_{11}}\right).
\end{equation}
Then according to \eqref{Scb1}, we have
\begin{equation} \label{ImS2}
ImS= Im \int p_r dr.
\end{equation}
Insert the components of metric \eqref{line element of the general Kerr-like metric in DFT} into \eqref{pr} and \eqref{ImS2}, we can obtain the explicit formula for $ImS$. Since it is too lengthy, we do not list the result. Finally, from \eqref{TomegaS}, we can obtain the Hawking temperature for the Kerr-like metric in $4D$ DFT, which is obtained in section 3. In fact, if we consider the non-rotating black hole, the result for the 'Schwarzschild-like' metric obtained in \cite{YL2022} can be obtained as a special case.

\section{Conclusions and Discussions}
In recent years, a significant advancement in string theory is the development of double field theory (DFT) \cite{Siegel19931,Siegel19932,D1990}. DFT aims to provide a manifestly duality-invariant description of supergravity. By introducing dual coordinates $\tilde{x}_{\mu}$, DFT allows fields and gauge parameters to depend on the $2D$ coordinates $X^M=(x^{\mu}, \tilde{x}_{\mu})$, forming an $O(D,D)$ vector. If we impose the "section condition" or "strong constraint," restricting fields and gauge parameters to depend on at most half of the coordinates, the action $S_{NSNS}$ can be derived from the action of DFT. Thus, DFT offers a natural framework for exploring T-duality-inspired possibilities beyond supergravity.

The study of black hole (BH) solutions in string theory is a dynamic area of research. A notable achievement of superstring theory is the correspondence between the microscopic state counting for the 5-dimensional BPS black hole solution of the string effective action and the Bekenstein-Hawking entropy \cite{SW1996,Z2023}. Subsequently, numerous efforts have been made to extend these results to other solutions in string theory \cite{HMS1996,S2008,CDKL20071,CDKL20072,CDKL2008,CM2009,KLL1999,KL2005}. The first black hole solution within double field theory was presented in \cite{KPS2017}, where the authors investigated a general static, asymptotically flat, spherically symmetric solution in $D=4$, sourced by the stringy energy–momentum tensor. This solution matches the known vacuum geometry outside a finite radius from the center, with parameters expressed as volume integrals of the interior stringy energy–momentum tensor. Additionally, the authors discuss the energy conditions of the black hole metric \cite{ACP2018}.

However, the metric proposed in \cite{KPS2017,ACP2018} is a 'Schwarzschild-like' metric in $D=4$ DFT. In this paper, we extend this metric to the rotating case, i.e., a 'Kerr-like' metric. Our investigation yields two main results. First, we derive the Kerr-like metric in $4D$ double field theory in both the Einstein frame and the string frame, following a methodology akin to deriving the Kerr metric in General Relativity. We complexify the relevant physical quantities in this metric. We discuss the conditions which $H$-flux and dilaton should satisfy as well. Second, we explore the thermodynamics and Hawking radiation of the Kerr-like metric obtained. We employ the 'covariant phase space' approach \cite{V.L and R.M.W 1994,R.M.W 1993,J.L and R.M.W 1990}, which reformulates the first law of black hole thermodynamics into a 'differential' form. We recover the usual integrated form of the first law of black hole thermodynamics \cite{ASA and CDAB 2017}. Moreover, we provide explicit formulas for the first law of black hole thermodynamics and the Hawking radiation temperature for the 'Kerr-like' metric in $4D$ DFT. Notably, the results for the 'Schwarzschild-like' metric are regained in the non-rotating case.

In conclusion, we offer a brief outlook on potential future developments. Firstly, we aim to derive additional black hole metrics within $4D$ double field theory. Secondly, we will explore whether the metrics in $D=4$ DFT satisfy the CA-duality ('complexity=action'). Lastly, we will investigate the disparities between solutions in $4D$ double field theory and General Relativity, with hopes of identifying observable differences in future astronomical observations.

\paragraph{Acknowledgements} 
YL was supported by an STFC studentship. For the purpose of open access, the authors have applied a CC BY public copyright licence to any Author Accepted Manuscript version arising.

\appendix

\section{Kerr-Newman metric in General Relativity}
In this appendix, we will review how to derive the Kerr-Newman metric in general relativity. We first write Reissner-Nordstroem metric for an electrically charged, static, and spherically symmetric body
\begin{equation} \label{eq: metric RN}
    ds^2_{RN}=-\left(1-\frac{2M}{r}+\frac{Q^2}{r^2}\right)dt^2+\left(1-\frac{2M}{r}+\frac{Q^2}{r^2}\right)^{-1}dr^2+r^2d\Omega^2,
\end{equation} 
where $d\Omega^2=d\theta^2+\sin{\theta}^2d\varphi^2$.
We set:
\begin{equation} \label{du in RN}
    du=-dt+\left(1-\frac{2M}{r}+\frac{Q^2}{r^2}\right)^{-1}dr.
\end{equation}
Then this metric can be rewritten as:
\begin{equation} \label{metric RN u}
    ds^2=-\left(1-\frac{2M}{r}+\frac{Q^2}{r^2}\right)du^2+2dudr+r^2d\Omega^2.
\end{equation}
Because $dr^2$ is eliminated, the metric can now be expressed in terms of a null tetrad $\{ l^{\mu},n^{\mu},m^{\mu},\bar{m}^{\mu} \} $. The vectors $l^{\mu},n^{\mu}$ are real while $m^{\mu},\bar{m}^{\mu}$ are complex conjugates; the only non-vanishing scalar products between tetrad vectors are:
\begin{equation} \label{l, n}
    l^{\mu} n_{\mu} = 1, n^{\mu} n_{\mu}=0,
\end{equation} 
and we can set 
\begin{equation} \label{lmu}
    l^{\mu}=\delta_1^{\mu}
\end{equation}
\begin{equation} \label{nmu}
n^{\mu}=\delta_0^{\mu}+C\delta_1^{\mu}, 
\end{equation}
where the $\delta$'s are the unit vector of cotangent space, and $C$ is an unknown function. From \eqref{l, n}, \eqref{lmu} and \eqref{nmu}, we have
\begin{equation} \label{C RN}
    C=-\frac{1}{2}\left(1-\frac{2M}{r}+\frac{Q^2}{r^2}\right).
\end{equation}
Combining \eqref{g up l n m} and $m^{\mu} \bar{m}_{\mu}=-1$, we have
\begin{equation} \label{m RN}
    m^{\mu}=\frac{1}{\sqrt{2}r}\left(\delta_2^{\mu}+\frac{i}{\sin{\theta}}\delta_3^{\mu}\right),
\end{equation}
\begin{equation} \label{bar m RN}
    \bar{m}_{\mu}=\frac{1}{\sqrt{2}r}\left(\delta_2^{\mu}-\frac{i}{\sin{\theta}}\delta_3^{\mu}\right).
\end{equation}
We need to do complexification and write the expression as:
\begin{equation} \label{r RN complexification }
    dr'=dr-ia \sin \theta d\theta,
\end{equation}
\begin{equation} \label{u RN complexification}
    du'=du+ia \sin\theta d\theta,
\end{equation}
Then based on transformations \eqref{r RN complexification } and \eqref{u RN complexification}, we can obtain
\begin{equation} \label{A RN complexification}
    1-\frac{2M}{r}+\frac{Q^2}{r^2} \rightarrow 1-\frac{2Mr-Q^2}{r'^2+a^2 \cos^2\theta },
\end{equation}
\begin{equation} \label{r2 RN complexification}
    r^2 \rightarrow r^2+a^2 \cos^2\theta,
\end{equation}
so
\begin{equation} \label{l RN complexification}
    l^{\mu}=\delta_1^{\mu},
\end{equation}
\begin{equation} \label{n RN complexification}
    n^{\mu}=\delta_0^{\mu}-\frac{1}{2}\left(1-\frac{2Mr-Q^2}{r'^2+a^2 \cos^2\theta }\right) \delta_1^{\mu},
\end{equation}
\begin{equation} \label{m RN complexification}
    m^{\mu}=\frac{1}{\sqrt{2}(r'^2+a^2\cos^2\theta) } \left(\delta_2^{\mu}-ia\sin\theta \delta_1^{\mu}+ia\sin\theta \delta_0^{\mu}+\frac{i}{\sin\theta} \theta_3^{\mu}\right).
\end{equation}
According to \eqref{g up l n m}, we can get:
\begin{equation} \label{g up RN complexification l n m}
    g^{\mu \nu}=\left[  \begin{array}{cccc}
\frac{a^2 \sin^2\theta}{\rho^2}  &  -\frac{r'^2+a^2}{\rho^2} & 0  &  \frac{a}{\rho^2}  \\
   -\frac{r'^2+a^2}{\rho^2}   &  \frac{r'^2+a^2-2Mr+Q^2}{\rho^2} & 0  &  -\frac{a}{\rho^2} \\
    0  &  0 & \frac{1}{\rho^2}  &  0 \\
  \frac{a}{\rho^2}    & -\frac{a}{\rho^2}  &  0 & \frac{1}{\rho^2 \sin^2\theta}
    \end{array}   \right],
\end{equation} 
and $\rho^2= r'^2+a^2\cos^2\theta$.
The inverse of \eqref{g up RN complexification l n m} is
\begin{equation} \label{g down RN complexification l n m}
    g_{\mu \nu}=\left[  \begin{array}{cccc}
-\left(1-\frac{2Mr-Q^2}{\rho^2}\right) &  -1 & 0  &  -\frac{a\sin^2\theta(2Mr-Q^2)}{\rho^2}  \\
   -1   &  0 & 0  & a\sin^2\theta \\
    0  &  0 & \rho^2  &  0 \\
  -\frac{a\sin^2\theta(2Mr-Q^2)}{\rho^2}    & a\sin^2\theta  &  0 & \sin^2\theta\left(r^2+a^2+a^2\sin^2\theta \frac{2Mr-Q^2}{\rho^2}\right)
    \end{array}   \right].
\end{equation}
We can use the following transformation to turn back to coordinates $(t,r,\theta, \phi )$
 \begin{equation} \label{transfor back RN}
      \left[\begin{array}{c}
         dt \\
         dr \\
         d\theta\\
         d\varphi
    \end{array}
    \right]=\left[ \begin{array}{cccc}
        1 & -\frac{r^2+a^2}{r^2+a^2-2Mr+Q^2} & 0 & 0 \\
        0 & 1 & 0 & 0 \\
        0 & 0 & 1 & 0\\
        0 & -\frac{a}{r^2+a^2-2Mr+Q^2} & 0 & 1
    \end{array}
    \right]
    \left[ \begin{array}{c}
        du \\
        dr \\
       d\theta \\
       d\varphi
    \end{array}\right],
 \end{equation}
After transforming the coordinates, we will find that $(0,1)$,$(1,0)$,$(1,3)$ and $(3,1)$ terms in following metric is vanishing \cite{carter 1968}. Then we can get the metric:
\begin{equation} \label{KN}
    g_{\mu \nu}=\left[  \begin{array}{cccc}
-\left(1-\frac{2Mr-Q^2}{\rho^2}\right) &  0 & 0  &  -\frac{a\sin^2\theta(2Mr-Q^2)}{\rho^2}  \\
   0 & \frac{\rho^2}{r^2+a^2-2Mr+Q^2} & 0  & 0 \\
    0  &  0 & \rho^2  &  0 \\
  -\frac{a\sin^2\theta(2Mr-Q^2)}{\rho^2}    & 0 &  0 & \sin^2\theta \left(r^2+a^2+a^2\sin^2\theta \frac{2Mr-Q^2}{\rho^2}\right)
    \end{array}   \right].
\end{equation}
More details can be found in \cite{T.A and E.T.N 2016}.

\section{Christoffel symbols of new metric}
We give the non-vanishing Christoffel symbols
\begin{equation}
    \begin{split}
        \Gamma_{tt}^{r}&=\frac{\hat{A}(\hat{C}+\psi^2\sin^2\theta)}{2\hat{C}}\left(\frac{\partial \hat{A}}{\partial r}+2\hat{A}\frac{\partial \hat{\phi}}{\partial r}\right)\\
        \Gamma^{\theta}_{tt}&=\frac{\hat{A}}{2\hat{C}}\left(\frac{\partial \hat{A}}{\partial r}+2\hat{A}\frac{\partial \hat{\phi}}{\partial r}\right)\\
        \Gamma^t_{tr}=\Gamma^t_{rt}&=\frac{\psi^2\sin^2\theta(\frac{\partial \hat{A}}{\partial r}+2\frac{\partial \hat{\phi}}{\partial r})+\hat{C}(\frac{\partial \hat{A}}{\partial r}+2\hat{A}\frac{\partial \hat{\phi}}{\partial r})}{2(\hat{A}\hat{C}+\psi^2\sin^2\theta)}\\
        \Gamma^{\varphi}_{tr}=\Gamma^{\varphi}_{rt}&=\frac{\psi \frac{\partial A}{\partial r}}{2(\hat{A}\hat{C}+\psi^2\sin^2\theta)}\\
        \Gamma^t_{t\theta}=\Gamma^t_{\theta t}&=\frac{\psi^2\hat{A}^2\sin(2\theta)+\hat{C}\frac{\partial \hat{A}}{\partial \theta}+\hat{A}(-4\psi^2\cos\theta\sin\theta+2\hat{C}\frac{\partial \hat{\phi}}{\partial \theta})}{2(\hat{A}\hat{C}+\psi^2\sin\theta)}\\
        &+\frac{\psi^2\sin\theta(2\cos\theta+\sin\theta\frac{\partial \hat{A}}{\partial \theta}+2\sin\theta\frac{\partial \hat{\phi}}{\partial \theta})}{2(\hat{A}\hat{C}+\psi^2\sin\theta)}\\
        \Gamma^{\varphi}_{t\theta}=\Gamma^{\varphi}_{\theta t}&=\frac{\psi\csc^2\theta(-2\hat{A}\cot\theta+2\Hat{A}^2\cot\theta+\frac{\partial \hat{A}}{\partial \theta})}{2(\psi^2+\hat{A}\csc^2\theta\hat{C})}\\
        \Gamma^{\theta}_{t\varphi}=\Gamma^{\theta}_{\varphi t}&=-\frac{\psi\hat{A}\sin^2\theta(\hat{C}+\psi^2\sin^2\theta)(\frac{\partial \hat{A}}{\partial r}+2(-1+\hat{A})\frac{\partial \Hat{\phi}}{\partial r})}{2\hat{C}}\\
        \Gamma^r_{rr}&=-\frac{\psi^2\hat{A}\sin^2\theta\frac{\partial \hat{C}}{\partial r}+\hat{C}^2(\frac{\partial \hat{A}}{\partial r}-2\hat{A}\frac{\partial \hat{\phi}}{\partial r})+\psi^2\hat{C}\sin^2\theta(\frac{\partial \hat{A}}{\partial r}-2\hat{A}\frac{\partial \hat{\phi}}{\partial r})}{2\hat{A}\hat{C}(\hat{C}+\psi^2\sin^2\theta)}\\
        \Gamma^{\theta}_{rr}&=\frac{\hat{C}(\hat{C}+\psi^2\sin^2\theta)\frac{\partial \hat{A}}{\partial \theta}-\hat{A}\sin^2\theta\frac{\partial \hat{C}}{\partial \theta}-2\hat{C}^2\hat{A}\frac{\partial \hat{\phi}}{\partial \theta}}{2\hat{A}\hat{C}(\hat{C}+\psi^2\sin^2\theta)^2}
        +\frac{\psi^2(\sin 2\theta-2\sin^2\theta\frac{\partial \hat{\phi}}{\partial \theta})00}{2\hat{A}\hat{C}(\hat{C}+\psi^2\sin^2\theta)^2}\\
        \Gamma^r_{r\theta}=\Gamma^r_{\theta r}&=\frac{1}{2}\left(-\frac{\partial \hat{A}}{\partial \theta}\frac{1}{\hat{A}}+\frac{\psi^2\sin^2\theta\frac{\partial \hat{C}}{\partial \theta}+2\hat{C}^2\frac{\partial \hat{\phi}}{\partial \theta}+2\psi^2\hat{C}\sin\theta(-\cos\theta+\sin\theta\frac{\partial \hat{\phi}}{\partial \theta})}{\hat{C}(\hat{C}+\psi^2\sin^2\theta)} \right)\\
        \Gamma^{\theta}_{r\theta}=\Gamma^{\theta}_{\theta r}&=-\frac{1}{2}\frac{\partial \hat{A}}{\partial r}\frac{1}{\hat{A}}+\frac{1}{2}\frac{\partial \hat{C}}{\partial r}\frac{1}{\hat{C}}+\frac{\partial \hat{\phi}}{\partial r}\\
        \Gamma^t_{t\varphi}=\Gamma^t_{\varphi t}&=-\frac{\psi\sin^2\theta(\hat{C}\frac{\partial\hat{A}}{\partial r}+\psi^2\sin^2\theta\frac{\partial \hat{A}}{\partial r}-(-1+\hat{A}\frac{\partial \hat{C}}{\partial r}))}{2\hat{A}\hat{C}+\psi^2\sin^2\theta}\\
        \Gamma^{\varphi}_{r\varphi}=\Gamma^{\varphi}_{\varphi t}&=\frac{-\psi^2\sin^2\theta\frac{\partial \hat{A}}{\partial r}+2\psi^2\sin^2\theta\frac{\partial \hat{\phi}}{\partial r}+\hat{A}(\frac{\partial \hat{C}}{\partial r}+2\hat{C}\frac{\partial\hat{\phi}}{\partial r})}{2\hat{A}\hat{C}+\psi^2\sin^2\theta}\\
        \Gamma^r_{\theta\theta}&=\frac{(\hat{C}+\psi^2\sin^2\theta)(-\hat{A}\frac{\partial \hat{C}}{\partial r}+\hat{C}(\frac{\partial \hat{A}}{\partial r}-2\hat{A}\frac{\partial \hat{\phi}}{\partial r}))}{2\hat{A}\hat{C}}\\
        \Gamma^{\theta}_{\theta\theta}&=-\frac{1}{2}\frac{\partial \hat{A}}{\partial \theta}\frac{1}{\hat{A}}+\frac{1}{2}\frac{\partial \hat{C}}{\partial \theta}\frac{1}{\hat{C}}+\frac{\partial \hat{\phi}}{\partial \theta}\\
    \end{split}
\end{equation}
\begin{equation}
    \begin{split}
        \Gamma^{t}_{\theta\varphi}=\Gamma^t_{\varphi\theta}&=-\frac{\psi\sin^2\theta(2\psi^2\sin2\theta+\psi^2\hat{A}\sin2\theta+\hat{C}\frac{\partial\hat{A}}{\partial \theta}+\psi^2\sin^2\theta\frac{\partial \hat{A}}{\partial \theta}+\frac{\partial \hat{C}}{\partial \theta})}{2(\hat{A}\hat{C}+\psi^2\sin^2\theta)}\\
        &+\frac{\psi\sin^2\hat{A}(3\psi^2\sin2\theta+\frac{\partial \hat{C}}{\partial \theta})}{2(\hat{A}\hat{C}+\psi^2\sin^2\theta)}\\
\Gamma^{\varphi}_{\theta\varphi}=\Gamma^{\varphi}_{\varphi\theta}&=\frac{-2\psi^2\hat{A}^2\cot\theta+\psi^2(-\frac{\partial \hat{A}}{\partial \theta}+2(\cot\theta+\frac{\partial \hat{\phi}}{\partial \theta}))}{2(\psi^2+\hat{A}\csc^2\theta\hat{C})}\\
&+\frac{\hat{A}(4\psi^2\cot\theta+\csc^2\theta\frac{\partial \hat{C}}{\partial \theta}+2\csc^2\theta\hat{C}\frac{\partial \hat{\phi}}{\partial\theta})}{2(\psi^2+\hat{A}\csc^2\theta\hat{C})}\\
        \Gamma^{r}_{\varphi\varphi}&=\frac{\hat{A}\sin^2\theta(\hat{C}+\psi^2\sin^2\theta)(\psi^2\sin^2\theta\frac{\partial \hat{A}}{\partial r}-\frac{\partial\hat{C}}{\partial r}-2(\hat{C}-\psi^2(-2+\hat{A})\sin^2\theta)\frac{\partial \hat{\phi}}{\partial r})}{2\hat{C}}\\
        \Gamma^{\theta}_{\varphi\varphi}&=\frac{\hat{A}\sin^2\theta(-8\psi^2\cos\theta\sin\theta+\psi^2\sin^2\theta\frac{\partial \hat{A}}{\partial \theta}-\frac{\partial\hat{C}}{\partial \theta}-4\psi^2\sin^2\frac{\partial \hat{\phi}}{\partial \theta}+2\psi^2\hat{A}\sin\theta(2\cos\theta+\sin\theta\frac{\partial \hat{\phi}}{\partial \theta}))}{2\hat{C}}\\
        &+\frac{\hat{A}\sin\theta(-2\hat{C}(\cos\theta+\sin\theta\frac{\partial \hat{\phi}}{\partial \theta}))}{2\hat{C}}
    \end{split}
\end{equation}


\begin{thebibliography}{99}

\bibitem{KPS2017}
S.M. Ko, J.H. Park, M. Suh, \emph{JCAP} {06} (2017)002.

\bibitem{Siegel19931}
W. Siegel, Phys.Rev.D \textbf{47} (1993), 5453

\bibitem{Siegel19932}
W. Siegel, \emph{Phys. Rev. D} {48} (1993)2826.

\bibitem{D1990}
M.J. Duff, \emph{Nucl. Phys. B} {335} (1990) 610.

\bibitem{ASA and CDAB 2017}
Alex S Arvanitakis and Chris D A Blair, \emph{Class. Quantum Grav.} {34} (2017) 055001.  

\bibitem{SW1996}
A. Strominger and C. Vafa, \emph{Phys. Lett. B} {379} (1996) 99.

\bibitem{Z2023}
M. Zatti, \emph{JHEP} {11} (2023) 185.

\bibitem{HMS1996}
G.T. Horowitz, J.M. Maldacena and A. Strominger, \emph{Phys. Lett. B} {383} (1996) 151.

\bibitem{S2008}
A. Sen, \emph{Gen. Rel. Grav.} {40} (2008) 2249.

\bibitem{CDKL20071}
A. Castro, J.L. Davis, P. Kraus and F. Larsen, \emph{JHEP} {06} (2007) 007.

\bibitem{CDKL20072}
A. Castro, J.L. Davis, P. Kraus and F. Larsen, \emph{JHEP} {09} (2007) 003.

\bibitem{CDKL2008}
A. Castro, J.L. Davis, P. Kraus and F. Larsen, \emph{Int. J. Mod. Phys. A} {23} (2008) 613.

\bibitem{CM2009}
A. Castro and S. Murthy, \emph{JHEP} {06} (2009) 024.

\bibitem{KLL1999}
D. Kutasov, F. Larsen and R.G. Leigh, \emph{Nucl. Phys. B} {550} (1999) 183.

\bibitem{KL2005}
P. Kraus and F. Larsen, \emph{JHEP} {09} (2005) 034.

\bibitem{ACP2018}
Stephen Angus, Kyoungho Cho, Jeong-Hyuck Park, \emph{EPJC} (2018) 78:500.

\bibitem{V.L and R.M.W 1994}
Vivek lyer and Robert M. Wald, \emph{Phys. Rev. D} {50} 846-64 

\bibitem{R.M.W 1993}
Robert M.Wald,  \emph{Phys. Rev. D} {48} 3427-31 

\bibitem{J.L and R.M.W 1990}
Joohan Lee and Robert M.  \emph{J.Math. Phys.} {31} 725-43 


\bibitem{RTF1980}
V. C. Rubin, N. Thonnard and W. K. Ford, Jr., \emph{Astrophys. J.} {238} (1980)471.

\bibitem{M1983}
M. Milgrom, \emph{Astrophys. J.} {270} (1983)365.




\bibitem{HZ20091}
C. Hull and B. Zwiebach, \emph{JHEP} {0909} (2009)099.

\bibitem{HZ20092}
C. Hull and B. Zwiebach, \emph{JHEP} {0909} (2009)090.

\bibitem{HZ20101}
O. Hohm, C. Hull and B. Zwiebach, \emph{JHEP} {1007} (2010)016.

\bibitem{HZ20102}
O. Hohm, C. Hull and B. Zwiebach, \emph{JHEP} {1008} (2010)008.



\bibitem{CP2022}
Kang-Sin Choi and Jeong-Hyuck Park, \emph{PRL} {129} (2022) 061603.

\bibitem{WY2014}
Houwen Wu and Haitang Yang, \emph{JCAP}, {\bf 07} (2014) 024.

\bibitem{T1990}
A.A. Tseytlin, \emph{Phys. Lett. B} {242} (1990) 163.


\bibitem{B2019}
Robert Brandenberger et al., \emph{Phys. Rev. D}, {\bf 99} (2019) 023531.

\bibitem{WeinbergGC}
S. Weinberg, Gravitation and Cosmology: Principles and Applications of the General Theory of Relativity (John Wiley and Sons, New York, 1972), ISBN 0471925675, 9780471925675.

\bibitem{YL2022}
Yang Liu, \emph{EPJC} (2022) 82:1054.

\bibitem{FM2000}
Laura Ferrarese and David Merritt, \emph{APJ} (2000), 539:L9–L12.

\bibitem{V2010}
Marta Volonteri, \emph{Astron Astrophys Rev} 18, 279–315 (2010).

\bibitem{IVH2020}
Kohei Inayoshi, Eli Visbal, and Zoltán Haiman, \emph{Annu. Rev. Astron. Astrophys.} 2020. 58:27–97.

\bibitem{JHP2019}
Jeong-Hyuck Park, \emph{PORCEEDINGS OF SCIENCE},  9 Apr 2019 

\bibitem{Parikh&Wilczek:2000}
M.K. Parikh, F. Wilczek, \emph{Phys. Rev. Lett.} {85}, 5042 (2000)

\bibitem{Johnson&March-Russell:2020}
G. Johnson, J. March-Russell, \emph{JHEP} {04}, 205 (2020)

\bibitem{Kraus&Wilczek:1995}
P. Kraus, F. Wilczek, \emph{Nucl. Phys. B} {433}, 403 (1995)

\bibitem{Parikh:2004}
M.K. Parikh, \emph{Int. J. Mod. Phys. D} {13}, 2351 (2004)

\bibitem{Parikh:2003}
M.K. Parikh, In: Proceedings of the 10th Marcel Grossmann Meeting (MG10, July 2026, Rio de Janeiro, Brazil, 2003) 

\bibitem{Vanzo&Acquaviva&Criscienzo:2011}
L.Vanzo, G. Acquaviva, R.Di Criscienzo, \emph{Class. Quantum Gravity} {28}, 183001 (2011)

\bibitem{Jiang&Wu&Cai:2006}
Qing-Quan Jiang, Shuang-Qing Wu, and Xu Cai, \emph{Phys. Rev. D} {73}, 064003 (2006)

\bibitem{T.A and E.T.N 2016}
Tim Adamoa and E. T. Newman, Scholarpedia 9(10):31791 arXiv:1410.6626v2 [gr-qc] 14 Nov 2016

\bibitem{carter 1968}
  B. Carter, \emph{Phys.Rev.} {174} (1968) 1559–1571.


\bibitem{CDAB 2016}
 C. D. A. Blair, \emph{JHEP} {04} (2016) 180,
[arXiv:1507.0754]. 

\bibitem{J.H.P S.J.R.W.R. and Y.S. 2015}
J.-H. Park, S.-J. Rey, W. Rim, and Y. Sakatani,  \emph{JHEP} {11} (2015) 131, 

\bibitem{U.N 2015}
 U. Naseer, \emph{JHEP} {10} (2015) 158,

\bibitem{DSP ETM and MJP 2011}
D. S. Berman, E. T. Musaev, and M. J. Perry, \emph{Phys.Lett. B} {706} (2011) 228–231,

\bibitem{OH and HS 2013}
O. Hohm and H. Samtleben,  \emph{Phys.Rev.D} {88} (2013) 085005, 

\bibitem{IR and RMW 1992}
I. Racz and R. M. Wald, \emph{Class. Quant. Grav.}
{9} (1992) 2643–2656. 

\bibitem{JL 2014}
J. Keir, \emph{Class. Quant. Grav.}  {31} (2014), no. 3 035014, 

\bibitem{KSC and JHP 2022}
Kang-Sin Choi and Jeong-Hyuck Park, \emph{Phys. Rev. Lett.} {129}, 061603 (2022) 

\bibitem{R.M.Wald 1984}
R.M.Wald, 1984, General Relativity, The University of Chicago Press Chicago and London 

\bibitem{K.S. STELLE 1997}
K.S. STELLE, [arXiv:hep-th/9701088].

\bibitem{EA and JC 2001}
Enrique Alvarez and Jorge Conde [arxiv.org/pdf/gr-qc/0111031]

\end{thebibliography}
\end{document}